\documentclass[aps,prb,twocolumn,superscriptaddress,10pt,showpacs]{revtex4-1}

\usepackage[utf8x]{inputenc}
\usepackage{graphicx}
\usepackage{amsmath,amsthm,amssymb}
\usepackage[dvipsnames]{xcolor}
\usepackage[colorlinks,linkcolor=blue,citecolor=blue,urlcolor=blue]{hyperref}

\usepackage{subfigure}

\graphicspath{{./figures/MF_GKBA_iTEBD/summary_figs/}}

\definecolor{darkpink}{rgb}{0.91, 0.33, 0.5}

\begin{document}

\title{Signatures of bosonic excitations in high-harmonic spectra of Mott insulators}
\author{Markus Lysne}
\affiliation{Department of Physics, University of Fribourg, 1700 Fribourg, Switzerland}
\author{Yuta Murakami}
\affiliation{Department of Physics, Tokyo Institute of Technology, Meguro, Tokyo 152-8551, Japan}
\author{Philipp Werner}
\affiliation{Department of Physics, University of Fribourg, 1700 Fribourg, Switzerland}

\date{\today}

\pacs{71.10.Fd}

\begin{abstract}
The high harmonic spectrum of the Mott insulating Hubbard model has recently been shown to exhibit plateau structures with cutoff energies determined by $n$th nearest neighbor doublon-holon recombination processes. The spectrum thus allows to extract the on-site repulsion $U$. Here, we consider generalizations of the single-band Hubbard model and discuss the signatures of bosonic excitations in high harmonic spectra. Specifically, we study an electron-plasmon model which captures the essential aspects of the dynamically screened Coulomb interaction in solids and a multi-orbital Hubbard model with Hund coupling which allows to analyze the effect of local spin excitations. 
For the electron-plasmon model, we show that the high harmonic spectrum can reveal information about the screened and bare onsite interaction, the boson frequency, as well as the relation between boson coupling strength and boson frequency. In the multi-orbital case, string states formed by local spin excitations result in an increase of the radiation intensity and cutoff energy associated with higher order recombination processes.  
\end{abstract}

\maketitle

\section{Introduction}
\label{sec:introduction}

High harmonic generation (HHG) is a highly non-linear process in which a laser field with a given fundamental frequency $\Omega$ generates ``overtones" in the emitted radiation, at multiples of the fundamental frequency.\cite{Corkum1993PRL, Lewenstein_1994,  Krausz_2009, Ghimire_2011, Ghimire_2018} HHG in atomic and molecular gases has been studied for decades\cite{Corkum1993PRL, Lewenstein_1994, Krausz_2009}, but the topic has gained renewed interest in recent years due to applications in condensed matter.\cite{Ghimire_2011, Schubert2014, Luu_2015,Vampa2015Nature,Langer2016Nature,Hohenleutner2015Nature,Ndabashimiye2016,Liu2017,You2016,Yoshikawa2017Science,Kaneshima2018, Ghimire_2018}  There are different motivations to study HHG in solids. 
On the one hand, a proper understanding of the physics underlying HHG may lead to table-top sources of high frequency radiation.\cite{Brabec_2000} On the other hand, HHG is also useful as a spectroscopic tool. The latter fact is exemplified by the recently demonstrated reconstruction of a material's band structure from its high harmonic spectrum \cite{Luu_2015,Vampa_2015b} and the measurement of the Berry curvature of a material.\cite{Luu_2018}

Both in atomic physics and in the condensed matter context the phenomenon of HHG has mostly been described in terms of single-particle pictures.\cite{Golde_2008, Ghimire_2011, Kemper2013NJP, Higuchi_2014, Vampa2014PRL, Vampa2015Nature, Wu2015,Tamaya2016,Luu2016,Otobe2016,Ikemachi2017,Osika2017,Hansen2017,Tancogne-Dejean2017b,Tancogne-Dejean2017,Ikemachi2018,Ikeda2018PRA} 
These studies revealed that the HHG in semiconductors originates from the intraband and interband dynamics of the excited electrons, where the latter is described by extensions of the successful three step model used in atomic systems.\cite{Vampa2014PRL,Ikemachi2017} Although these analyses provide an intuitive understanding of the basic mechanisms of HHG in semiconductors, we currently lack a detailed understanding of the effect of correlations, which may be essential for justifying the ultrafast dephasing conventionally used.\cite{Orlando2018,Orlando2019} Furthermore, HHG spectra from different types of condensed matter systems, such as amorphous systems or liquids, have  recently been reported. \cite{You_2017, Luu_2018_2,Luu2018Amorphas, Chinzei_2020} These observations also raise questions about the role of correlations and the possibility of HHG from materials which are not band insulators, and motivate studies of strongly correlated systems. 
Several numerical simulations of HHG in Mott insulators as well and other strongly correlated systems have been conducted\cite{Silva_2018, Murakami_2018a, Murakami_2018b, Dejean_2018, Imai_2019} and it has been shown that in the strong-field regime, the characteristic features of the high harmonic spectrum can be understood by considering quasi-local processes. \cite{Murakami_2018a, Murakami_2018b}

Theoretical investigations in this field pose technical challenges since they require numerical techniques capable of treating strongly correlated systems and at the same time non-perturbative driving fields. Dynamical mean field theory (DMFT)\cite{Georges_1996} has become a standard tool for the study of strongly correlated electron systems in equilibrium, thanks to the development of powerful methods (impurity solvers) for the solution of the DMFT equations.\cite{Rubtsov_2005, Dai_2005, Werner_2006} Recently, several methods for solving the time-dependent impurity problem in nonequilibrium DMFT\cite{Aoki_2014} have been developed.\cite{Werner_2009, Eckstein_2010, Werner_2013, Murakami_2015} 
Some of these methods allow to simulate strong field physics in strongly correlated materials and allow to access long enough times that the high harmonic spectrum produced by a few-cycle electric field pulse can be computed. 

The aim of this work is to extend the previous Hubbard-model based analysis of HHG in Mott insulators \cite{Silva_2018, Murakami_2018a} to more complicated but realistic systems and to reveal various ways in which HHG can act as a spectroscopic tool. In particular, we will focus on models which admit bosonic excitations and discuss the resulting signatures in the high harmonic spectrum. Specifically, we will consider a model with a dynamically screened interaction which changes from a large bare value at frequencies much above some plasmon frequency to a reduced screened interaction in the static limit, and we will show that HHG allows one to reveal several characteristic energy scales of these systems. We will also discuss a multi-orbital Hubbard model in which the motion of charge carriers produces string states by creating local Hund excitations. The annihilation of these strings will be shown to enhance the high-energy radiation compared to the single-band Hubbard model. 

The paper is organized as follows. In Sec.~\ref{sec:method} we describe the DMFT method used and the models considered in our study. The results for the electron-plasmon model are presented in Sec.~\ref{sec:plasmon} and those for the two-orbital model in Sec.~\ref{sec:2orbital}. The conclusions of our study are summarized in Sec.~\ref{sec:conclusions}.

\section{Method} 
\label{sec:method}

\subsection{DMFT for a Bethe lattice with electric field}

We consider lattice models with a Hamiltonian of the general form $H_\text{latt}=\sum_i H_{\text{loc},i}+\sum_{\langle i,j\rangle}H_{\text{hop},i,j}$, where $H_{\text{loc},i}$ describes the interaction and chemical potential terms on site $i$ and $H_{\text{hop},i,j}$ the hopping between sites $i$ and $j$, which is diagonal in the spin and orbital indices $\sigma$ and $\alpha$. This lattice model is solved within the DMFT approximation\cite{Georges_1996} which maps the lattice problem onto a self-consistently determined quantum impurity model of the form $H_\text{imp}=H_\text{loc}+H_\text{bath}+H_\text{hyb}$. Here, $H_\text{loc}$ is the same local Hamiltonian as in the lattice model, $H_\text{bath}$ describes a bath of noninteracting electrons whose parameters are optimized to mimic the lattice environment, and $H_\text{hyb}$ describes the hybridization between the impurity and the bath. In an action formulation the bath is integrated out and replaced by a hybridization function $\Delta_{\alpha,\sigma}(t,t')$ which controls how electrons hop in and out of the impurity orbital.  
Assuming that the lattice self-energy is local, this hybridization function is optimized in such a way that the Green function of the impurity, $G_{\text{imp},\alpha,\sigma}(t,t')$, is the same as the local lattice Green function, $G_{\text{latt},i,i,\alpha,\sigma}(t,t')$. Since we use the nonquilibrium formalism, the time indices are defined on the L-shaped contour,\cite{Aoki_2014} which runs from time 0 to some time $t_\text{max}$ along the real-time axis, back to time zero, and then to time $-i\beta$ (with $\beta$ the inverse temperature) along the imaginary-time axis. The basic classes and routines for the handling nonequilibrium Green's functions, on which our simulations are built, have recently been published in Ref.~\onlinecite{Schuler_2019}.

We will consider a lattice with a semi-circular density of states (the infinitely connected Bethe lattice).
For this lattice, the self-consistency relation can be expressed in a simple form.\cite{Georges_1996} In equilibrium, it reads
\begin{equation}
\Delta_{\alpha,\sigma}(t,t')=v_\alpha G_{\text{imp},\alpha,\sigma}(t,t')v_\alpha,
\end{equation}  
with $v_\alpha$ corresponding to one quarter of the noninteracting bandwidth for band $\alpha$.

Since we are interested in effects induced by strong electric fields, let us briefly explain how the electric field enters into DMFT calculations with Bethe-type self-consistency (details can be found in Ref.~\onlinecite{Werner_2017b}). The idea is to distinguish hopping processes ``parallel to the field" and ``antiparallel to the field" and to add the corresponding Peierls phases to the hopping terms in the Bethe-lattice type self-consistency equation:
\begin{align}
\Delta_{\alpha,\sigma}(t,t')=&\frac{1}{2}\Big[v_\alpha e^{i\phi(t)}G_{\text{imp},\alpha,\sigma}(t,t')v_\alpha e^{-i\phi(t')}\nonumber\\
&+v_\alpha e^{-i\phi(t)}G_{\text{imp},\alpha,\sigma}(t,t')v_\alpha e^{i\phi(t')}\Big]\nonumber\\
& \equiv \Delta_{L,\alpha,\sigma}+\Delta_{R,\alpha,\sigma},
\end{align}
where $\phi(t)=-\int_0^t ds E(s)ea/\hbar c$ is the Peierls phase for the electric field with amplitude $E$ and $a$ is the lattice spacing. The factor $1/2$ is a convention used to recover the usual Bethe lattice self-consistency in the model without field. In this set-up the kinetic energy and current can be measured as follows:
\begin{align}
E_\text{kin}(t)&= \text{Re}[\Gamma_{L,\alpha,\sigma}(t) + \Gamma_{R,\alpha,\sigma}(t)],\\
j(t)&= \text{Im}[\Gamma_{L,\alpha,\sigma}(t) - \Gamma_{R,\alpha,\sigma}(t)],
\end{align}
with $\Gamma_{L/R}(t)=-i[G_\text{imp}*\Delta_{L/R}]^<(t,t)$. While a Bethe lattice with field may at first sight look suspicious, the above procedure is consistent with the spirit of DMFT, and as we will show below, it reproduces all the electric field induced features which have been previously discussed for a hypercubic lattice implementation.\cite{Murakami_2018a} As in the latter work, the HHG intensity is calculated as the square of the Fourier transform of the dipole acceleration $(d/dt)j(t)$, i.e., as $|\omega j(\omega)|^2$. 

The impurity problem will be solved by the non-crossing approximation NCA, which is the lowest order self-consistent expansion in the hybridization function $\Delta$.\cite{Keiter_1971,Eckstein_2010,Werner_2013} This method is numerically cheap and is expected to give qualitatively correct results in the Mott insulating regime where the local interaction term dominates the hybridization term. 

In the following subsections, we will describe in some more detail the two models which will be studied in this paper, namely a Holstein-Hubbard model representing electrons coupling to plasmons, and a two-orbital Hubbard model with Hund coupling. From now on we will set $v_\alpha=1$ (bare bandwidth $4$), i.e. we measure energies in units of $v_\alpha$ and time in units of $\frac{\hbar}{v_\alpha}$. The lattice spacing and $\hbar$ are set to unity.

\subsection{Holstein-Hubbard model}

In order to investigate the effects of an electron-plasmon coupling, we consider a Hubbard-Holstein model with a single orbital per site and a local Hamiltonian of the form
\begin{align}
H_\text{loc}=&U_\text{bare} n_\uparrow n_\downarrow -\mu (n_\uparrow +n_\downarrow)\nonumber\\
&+g(n_\uparrow +n_\downarrow-1)(b+b^\dagger) + \omega_0 b^\dagger b,
\end{align}
with $n_\sigma$ the density for spin $\sigma$, $\mu$ the chemical potential, and $U_\text{bare}$ the bare on-site repulsion.  The electrons are coupled via local density fluctuations to bosons with frequency $\omega_0$. The electron-boson coupling is $g$ and the boson creation operator is denoted by $b^\dagger$. 

\begin{figure}[t]
\centering
\includegraphics[width=\columnwidth]{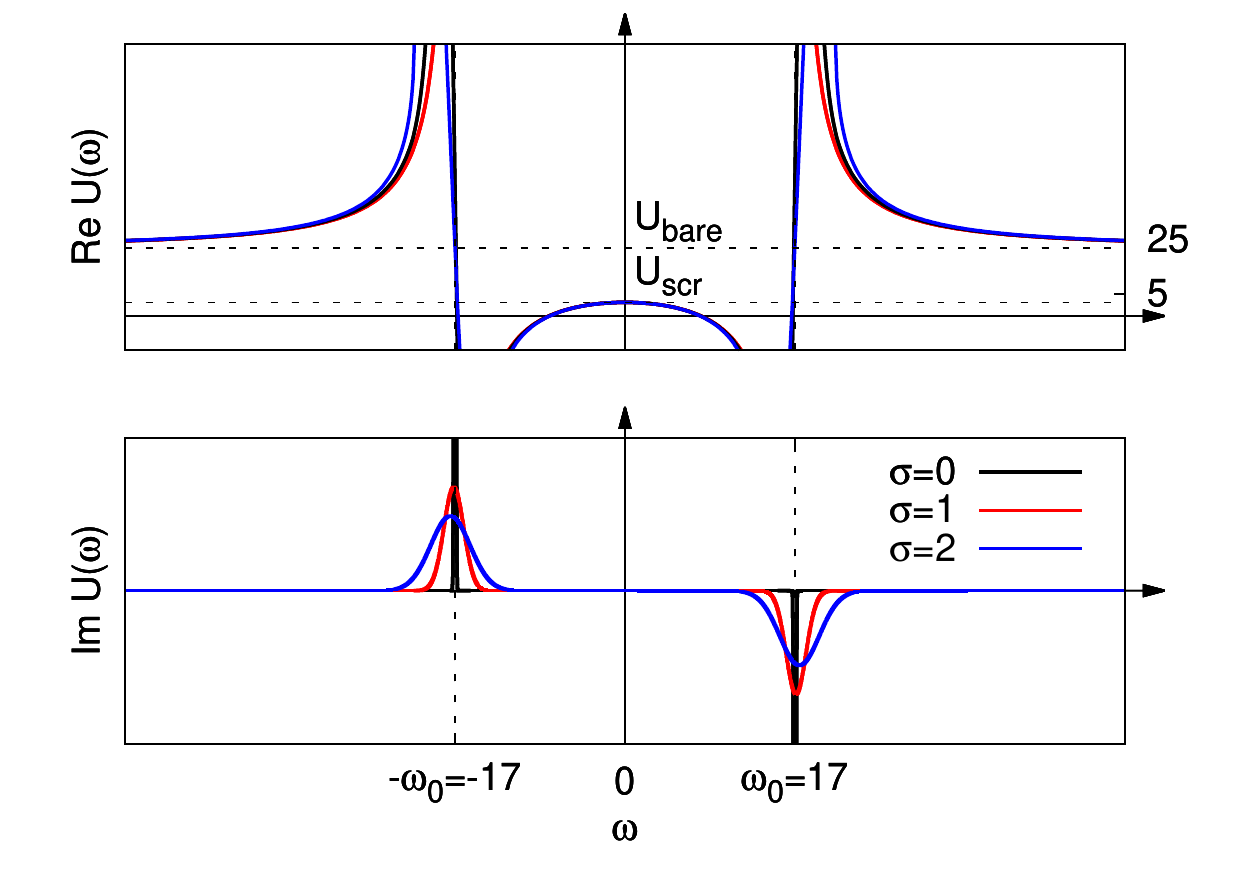}
\caption{Real and imaginary parts of the frequency dependent
 interaction, $U(\omega)$, for $\omega_0=17$, $U_\text{scr}=5$ and $\lambda=20$. The horizontal dashed lines are the asymptotic values for Re$U$ in the limit of $\omega \rightarrow \pm \infty$ and $\omega\rightarrow 0$. Besides the single boson case ($\sigma=0$) we also show the results for a distribution of bosons with a width defined by $\sigma$ (see Eq.~(\ref{multiboson})).
 }   
\label{fig:bosons2}
\end{figure}

\begin{figure}[t]
\centering
\includegraphics[width=\columnwidth, bb=0 10 385 260, clip=false]{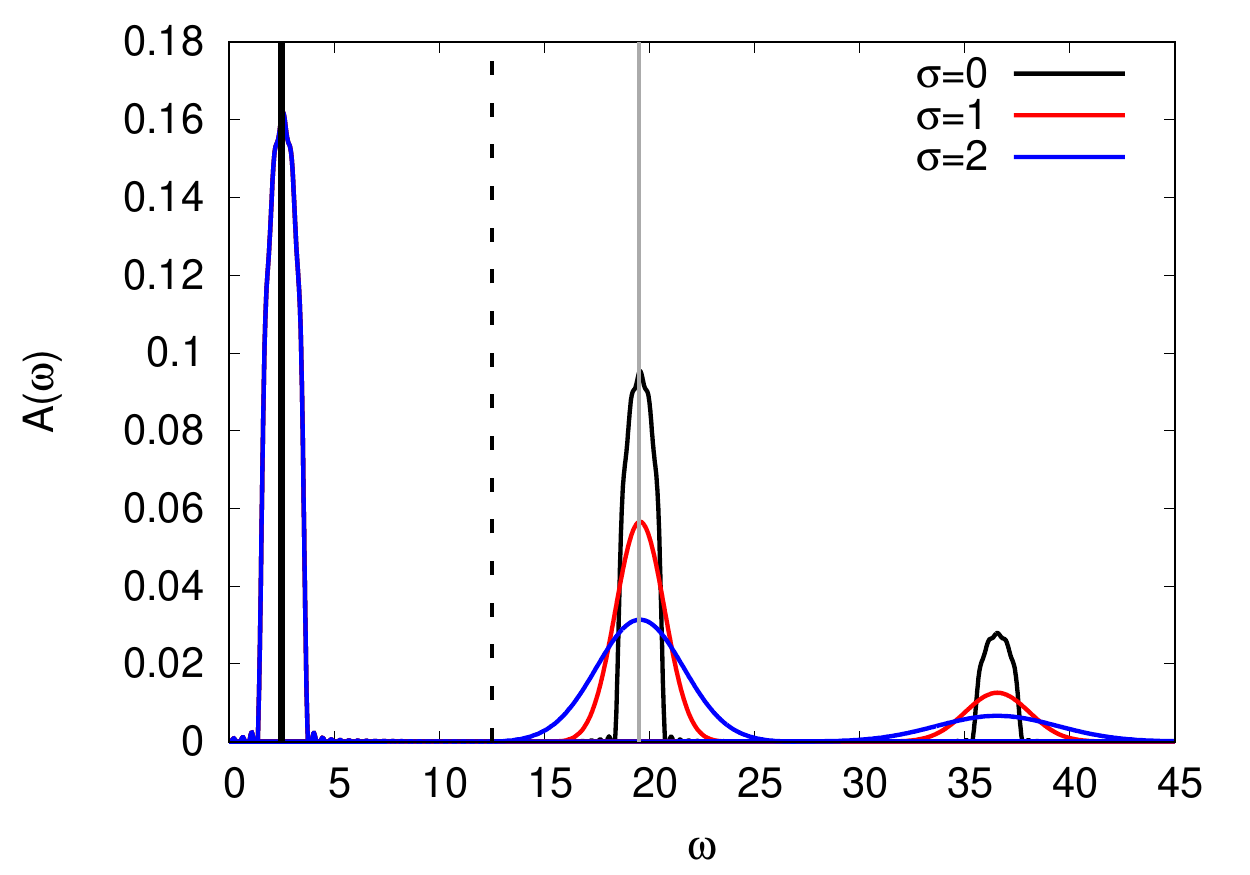}
\caption{Spectral functions for different boson distributions with $\sigma=0,1,2$. The parametes are $\omega_0=17$, $U_{\textrm{scr}}=5$, $\lambda=20$, and the inverse temperature is $\beta=5$. For $\sigma=1$ ($2$) we use $31$ ($61$) bosonic modes. The vertical lines show $U_\text{scr}/2$ (solid black) and $U_\text{bare}/2$ (dashed black) and the line $U_\text{scr}/2 + \omega_0$ (gray). For the main Hubbard band, all the data overlap. }   
\label{fig:spectral_function}
\end{figure}

In an action formulation, the bosons can be integrated out, which results in a retarded, or frequency dependent, effective interaction between the electrons.\cite{Altland_2010, Werner_2016, Murakami_2015} On the Matsubara axis, this frequency dependent interaction has the form 
\begin{equation} \label{I13}
    U(i\omega_n) = U_{\textrm{bare}} + \frac{2g^2 \omega_0}{(i\omega_n)^2-\omega_0^2}.
\end{equation}
Upon analytical continuation ($i\omega_n \rightarrow \omega + i0^{+}$), the real and imaginary parts become
\begin{equation} \label{I14}
\begin{aligned}
    &\textrm{ReU}(\omega) = U_{\textrm{bare}} + \frac{2g^2\omega_0}{\omega^2-\omega_0^2}, \\
    &\textrm{ImU}(\omega) =-g^2\pi(\delta(\omega-\omega_0)-\delta(\omega+\omega_0)).
\end{aligned}
\end{equation}
Hence, electrons oscillating with a frequency $\omega$ much higher than $\omega_0$ experience an effective interaction $U\approx U_{\textrm{bare}}$, whereas for $\omega \ll \omega_0$, the electrons experience $U\approx U_{\textrm{bare}}- \frac{2g^2}{\omega_0}\equiv U_{\textrm{scr}}$. In the following, we denote the difference between the bare and screened $U$ by $\lambda$ (=$\frac{2g^2}{\omega_0}$). 
The real and imaginary parts of $U(\omega)$ for a set of parameters corresponding to a large plasmon energy $\omega_0$ are shown by the black lines in Fig.~\ref{fig:bosons2}.\footnote{Our choice of parameters ($\text{bandwidth}=4$, $\omega_0=17$, $U_\text{scr}\approx 5$, $U_\text{bare}\approx 25$) may be considered representative of transition metal oxides.}
One notices a transition between the bare and screened regime with a pole-like structure around $\pm\omega_0$. In the spectral function of the Mott insulating Holstein-Hubbard model, these structures lead to plasmon satellite peaks which are split off from the Hubbard bands at $\pm U_\text{scr}/2$ by energies of $\pm n\omega_0$, see black spectrum in Fig.~\ref{fig:spectral_function}. In the equilibrium system at low temperature, only the high-energy sidebands are visible because they correspond (in the case of the upper band) to the insertion of an electron with simultaneous emission of a boson. (The analogous process with absorption of a boson is suppressed, because the bosonic system is in the ground state.)  

Sharp singularities as in the effective $U(\omega)$ for the Holstein-Hubbard model are not present in the downfolded effective interactions of realistic materials, obtained for example by the constrained random phase approximation.\cite{Ayrasetiawan_2004} In  realistic systems, the plasmon couples to single-particle excitations and the delta-function like structure of the plasmon in $\text{Im}U(\omega)$ becomes broadened (the pole-like structure in $\text{Re}U(\omega)$ becomes a smooth crossover from $U_\text{bare}$ to $U_\text{scr}$).\cite{Miyake_2008, Casula_2012, Werner_2015b} 

In order to model such a more realistic situation we can extend the single boson model to a model with a distribution of bosons, where the coupling constant for the boson with frequency $\omega_i$ obeys 
\begin{equation}
	\big( \frac{g_i}{\omega_i} \big )^2 = \big( \frac{g}{\omega_0} \big )^2 \frac{\exp{ (- \frac{(\omega_i-\omega_0)^2}{2\sigma^2} )} }{ \sum_i \exp{( - \frac{(\omega_i-\omega_0)^2}{2\sigma^2} )} }.
	\label{multiboson}
\end{equation}  
This choice of coupling constants ensures that the renormalized hopping (width of the main Hubbard bands near $\pm U_\text{scr}/2$) is the same in the single-boson and multi-boson case, since 
$t_{\textrm{eff}}=t \exp{(-\frac{g^2}{\omega_0^2})}$, where $t$ is the hopping parameter in the absence of electron boson coupling. \cite{Werner_2016}  
Furthermore, $\lambda$ stays the same which can be verified straightforwardly by computing $\sum_i \frac{2g_i^2}{\omega_i}$ with Eq.~\eqref{multiboson}.
The spectral functions for $\sigma=0,1,2$, $\omega_0=17$, $U_{\textrm{scr}}=5$ and $\lambda=20$ are shown in Fig.~\ref{fig:spectral_function} and the corresponding $U(\omega)$ are plotted in Fig.~\ref{fig:bosons2}. Here, we adjusted the number of bosonic modes $\omega_i$ such that the energy separation between neighboring modes is constant with spacing $0.2$, while the modes extend to a fixed $r\sigma$ of the Gaussian distribution, with $r$ some fixed real number.  Note that in the multi-boson case, the sidebands of the main Hubbard band get broadened, but the main Hubbard bands are left unaltered.

For the treatment of bosonic couplings in NCA, we use the procedure detailed in Ref.~\onlinecite{Werner_2013}, which has previously been used in nonequilibrium studies of electron-phonon problems.\cite{Werner_2013,Werner_2015a,Sayyad_2019} This method involves an approximation in the treatment of the boson couplings which is well justified in the limit of high boson frequency. It is thus particularly suited for the study of plasmon excitations, which have energies comparable to or larger than the bandwidth.  

We note in Fig.~\ref{fig:spectral_function} that in contrast to the energy scale $U_\text{scr}$, which fixes the position of the first (main) band, and $\omega_0$, which determines the energy splitting between sidebands, the energy scale $U_\text{bare}$ is not directly evident in the single-particle spectrum. Furthermore, the weight of the boson side-peaks monotonically decreases with increasing energy. 
Numerical evidence suggests that these are generic properties for systems with large $\omega_0 > g$. 

\begin{figure}[t]
\centering
\includegraphics[width=0.9\columnwidth, bb=5 10 385 260, clip=false]{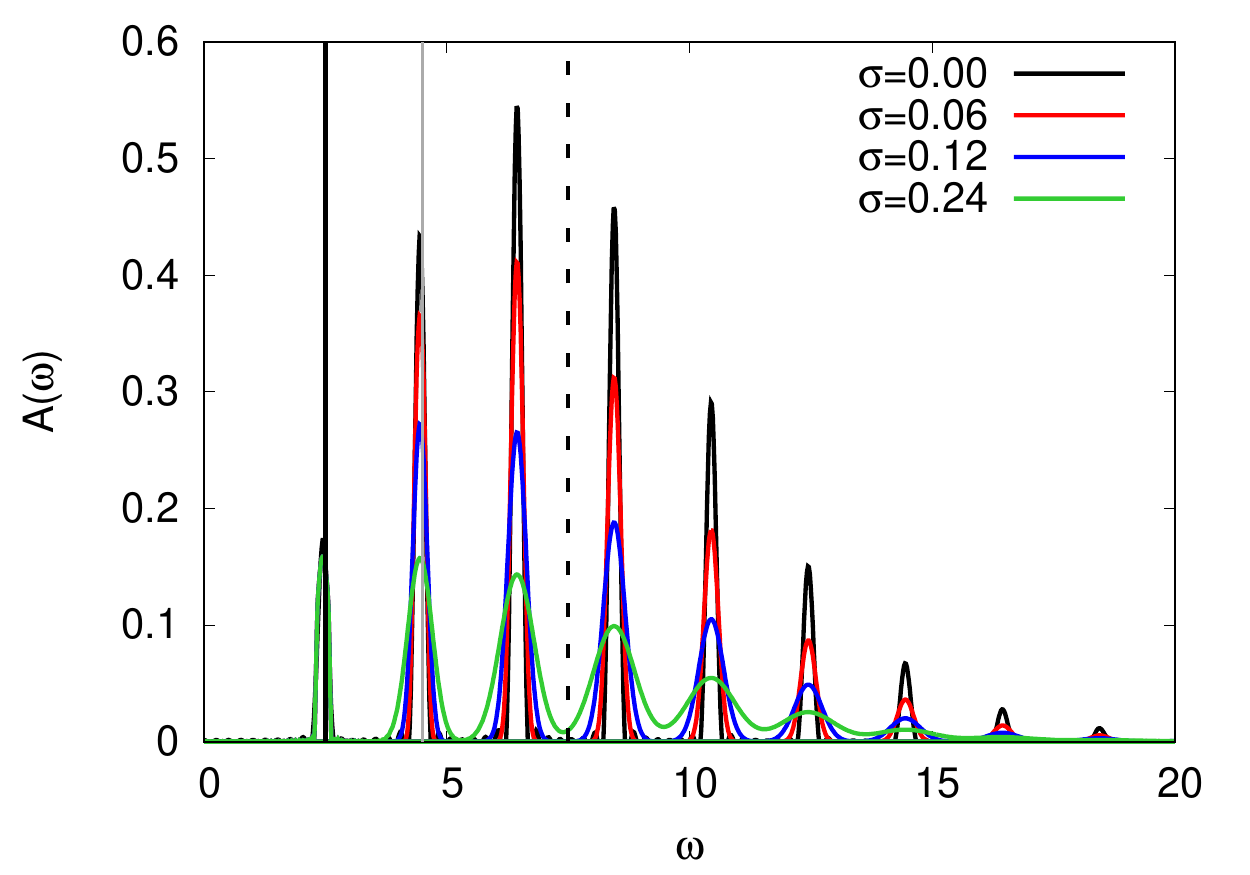}
\caption{Spectral functions for $\omega_0=2$, $U_{\textrm{scr}}=5$, $\lambda=10$, $\beta=5$, and different boson distributions parametrized by $\sigma$. The vertical lines indicate $U_{\text{scr}}/2$  (solid black), $U_{\text{bare}}/2$ (dashed black) and $U_\text{scr}/2+\omega_0$ (gray).}   
\label{fig:spectral_function_small_w0}
\end{figure}

A qualitatively different situation is encountered for $\omega_0 < g$, as illustrated in Fig.~\ref{fig:spectral_function_small_w0}, which shows the spectral function for $\omega_0=2$, $U_{\textrm{scr}}=5$ and $\lambda=10$. This parameter regime may be relevant for the description of sub-plasmons, which are collective excitations within a subset of orbitals. Here, the lowest peak in the upper Hubbard band is still at an energy close to $U_\text{scr}/2$, but an envelope drawn over the relatively tightly spaced subbands exhibits a peak around $U_{\textrm{bare}}/2$, so that the energy scale $U_\text{bare}$ manifests itself clearly in the single-particle spectrum. 

Based on the qualitative features of the spectral functions illustrated in Figs.~\ref{fig:spectral_function} and \ref{fig:spectral_function_small_w0} we can speculate how the 
HHG spectrum will look when plotted in the space of field strength $E_0$ and frequency $\omega$. An earlier analysis of the HHG response of the single-band Hubbard model in the Mott regime has found that the harmonic intensity is strong in a triangular region defined by $U-E_0 \lesssim \omega \lesssim U+E_0$, where $U$ is the on-site interaction of the Hubbard model and $E_0$ the amplitude of the AC electric field.\cite{Murakami_2018a} This suggests that the dominant contribution to the high harmonic emission can be attributed to the recombination of doublons and holons from nearest neighbor sites. At high field strength, higher order processes with associated cutoff energies $U+nE_0$ ($n>1$) could also be identified. The linear scaling of the cutoff as a function of field strength has become a hallmark of high harmonic generation in solids. \cite{Ghimire_2018, Wegener_2005} 

In the large-$\omega_0$ case ($\omega_0>g$) we expect to find a similar behavior with cutoff laws determined by $U_\text{scr}$. In addition, we may expect to see features in the HHG spectrum associated with transitions from plasmon side bands, i.e. the absorption of plasmons, once the field strength exceeds $\omega_0$ and a large number of plasmons is excited. In models  with $\omega_0<g$, where the screened and bare interaction determine the edge and the peak of the Hubbard band, respectively, we instead expect that the cutoff energies of the HHG plateaus will depend on the value of $E_0$ relative to the interaction strength. For $E_0\lesssim U_\text{scr}$, excitations between the lower and upper gap edge should govern the dynamics, and hence the first energy cutoff is expected to scale as $U_{\textrm{scr}}+E_0$. For $E_0\gtrsim U_\text{bare}$ transitions between the dominant peaks in the single-particle spectrum, located near $\pm U_{\textrm{bare}}/2$ will play a dominant role, so that the dominant cutoff may exhibit a scaling which is closer to $U_{\textrm{bare}}+E_0$ in the strong field regime. 

\begin{figure}[t]
\begin{center}
\includegraphics[angle=-0, width=0.49\columnwidth, bb=50 25 585 650, clip=true]{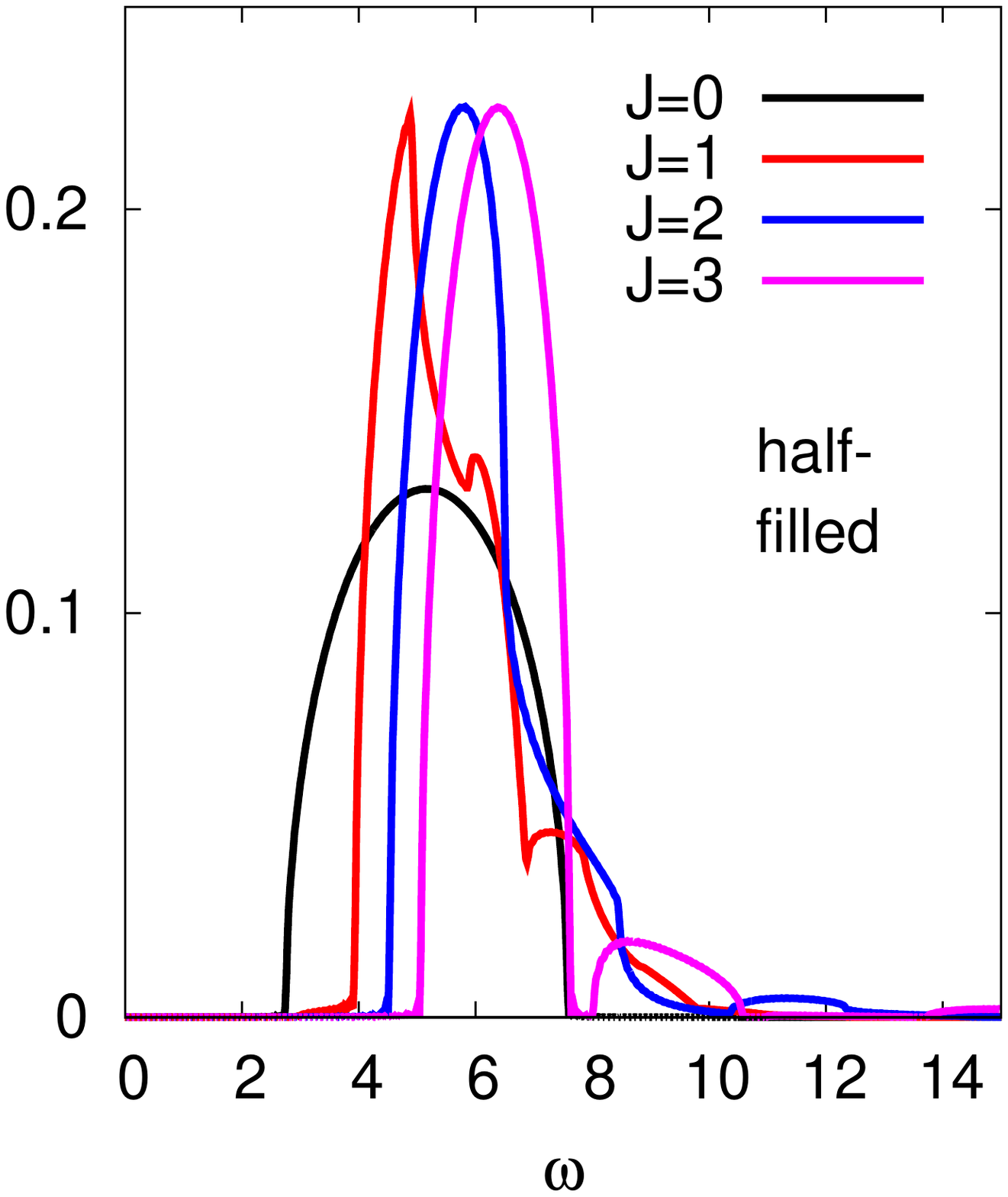}
\includegraphics[angle=-0, width=0.49\columnwidth, bb=50 25 585 650, clip=true]{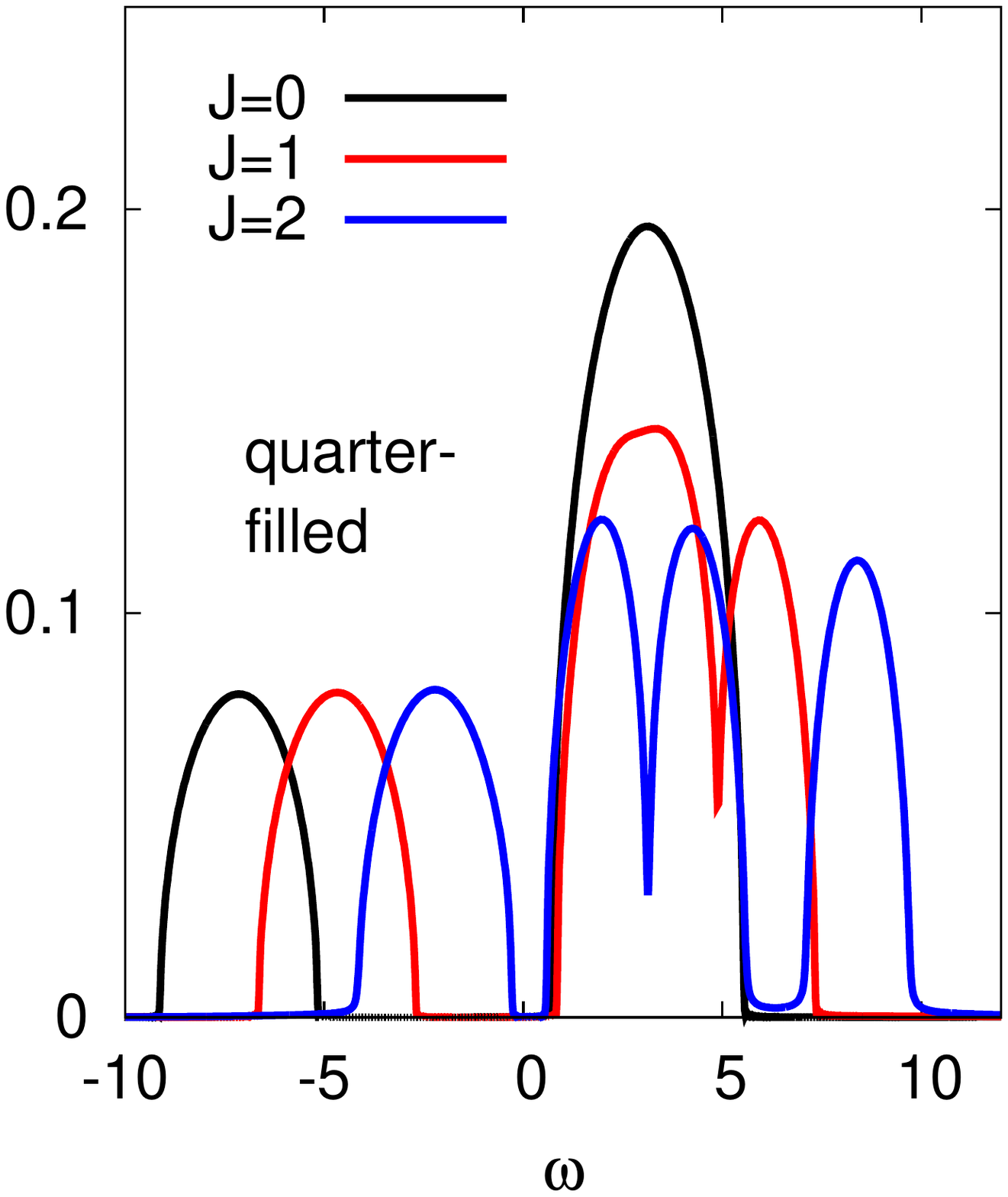}
\caption{Equilibrium spectral functions of the two-orbital model for $U=10$, inverse temperature $\beta=5$ and indicated values of $J$. The left panel is for the half-filled system and the right panel for the quarter-filled system.}
\label{fig:a}
\end{center}
\end{figure}

\subsection{Two-orbital model}
\label{sec:twoorbitaltheory}

For the two-orbital Hubbard model we consider a local Hamiltonian with density-density interactions of the form
\begin{align}
H_\text{loc}=&U \sum_{\alpha=1,2} n_{\alpha\uparrow} n_{\alpha\downarrow} + (U-2J) \sum_{\sigma} n_{1\sigma} n_{2\bar\sigma} \nonumber\\
&+ (U-3J) \sum_{\sigma} n_{1\sigma} n_{2\sigma},
\end{align}
with $U$ the intra-orbital repulsion and $J$ the Hund coupling. 

Low-temperature states of the half-filled system with $J>0$ are dominated by doubly occupied sites ($N=2$) in a high-spin configuration.  
If we denote the lowest energy state of $H_\text{loc}$ with occupation $N$ by $E_N$, then the Mott gap can be estimated as $\Delta_\text{Mott}(N)=E_{N+1}+E_{N-1}-2E_N$.\cite{Werner_2009} Thus, the Hubbard bands in the half-filled Mott state are expected near $\omega\approx \pm \frac{U+J}{2}$. 
The NCA spectral functions of half-filled systems are shown for $U=10$ and different $J$ in the left panel of Fig.~\ref{fig:a}. While the above argument explains the position of the main Hubbard bands, we also recognize a strong narrowing of these bands and the appearance of side-bands with increasing $J$. 

The satellites are associated with local spin excitations (Hund excitations) and the reduction of the kinetic energy is a consequence of strings of Hund excitations formed by charge carriers (``singlons" or ``triplons") moving in the half-filled Mott background. 
Looking at the equilibrium spectral functions shown in Fig.~\ref{fig:a} we anticipate that these Hund excitation processes can be clearly identified at large $J$, where the spectral function shows a well-defined satellite with an energy separation $J$. 

More specifically, the process contributing to this satellite is triplon creation plus hopping: $(\uparrow,\uparrow)_j (\downarrow,\downarrow)_{j+1} \rightarrow (\uparrow\downarrow,\uparrow)_j (\downarrow,\downarrow)_{j+1} \rightarrow (\downarrow,\uparrow)_j (\uparrow\downarrow,\downarrow)_{j+1}$.
The hopping in the last step costs an extra energy $(U-2J)-(U-3J)=J$, and in a background of high-spin states leads to a short string-like distortion. It is natural to assume that in the HHG spectrum, such processes affect in particular the cutoff energies of the second and higher plateaus, since these are associated with singlon-triplon recombination plus additional hoppings.

The quarter filled two-orbital model represents a different case from the half-filled model, which is evident by comparing the spectral functions in the right panel of Fig.~\ref{fig:a} with those in the left panel. Adding an electron to a singly occupied site creates doublon states with energies $U-3J$, $U-2J$ or $U$, so that the upper Hubbard band for large $J$ splits into three subbands. While triplons moving in a half-filled background can leave behind a string of excited doublon states, there is no such mechanism in the quarter filled case, where all the singlons have the same energy. We thus expect different $J$-related effects in the high-harmonic spectrum of the half-filled and quarter-filled model.

\section{Results}
\label{sec:results}

\subsection{Electron-plasmon model}
\label{sec:plasmon}

This section presents the results for the single-band electron-plasmon model. In the present study, we excite the system with a few-cycle electric field pulse. The form of the 10 cycle pulse with a central frequency of $\Omega=1$ and a $\sin^2$ envelope is shown in the inset of  Fig.~\ref{fig:E_vs_w_panel}. This set-up is different from Ref.~\onlinecite{Murakami_2018a}, which employed a Floquet DMFT formalism for time-periodic steady states. While the pulse protocol may lead to somewhat blurred high-harmonic features, it is more realistic from an experimental point of view. 

\subsubsection{Cut-off behavior for $\omega_0>g$ }

\begin{figure}[t]
  \centering
   \includegraphics[angle=0, width=\columnwidth, width=\columnwidth, bb=90 70 750 530, clip=true]{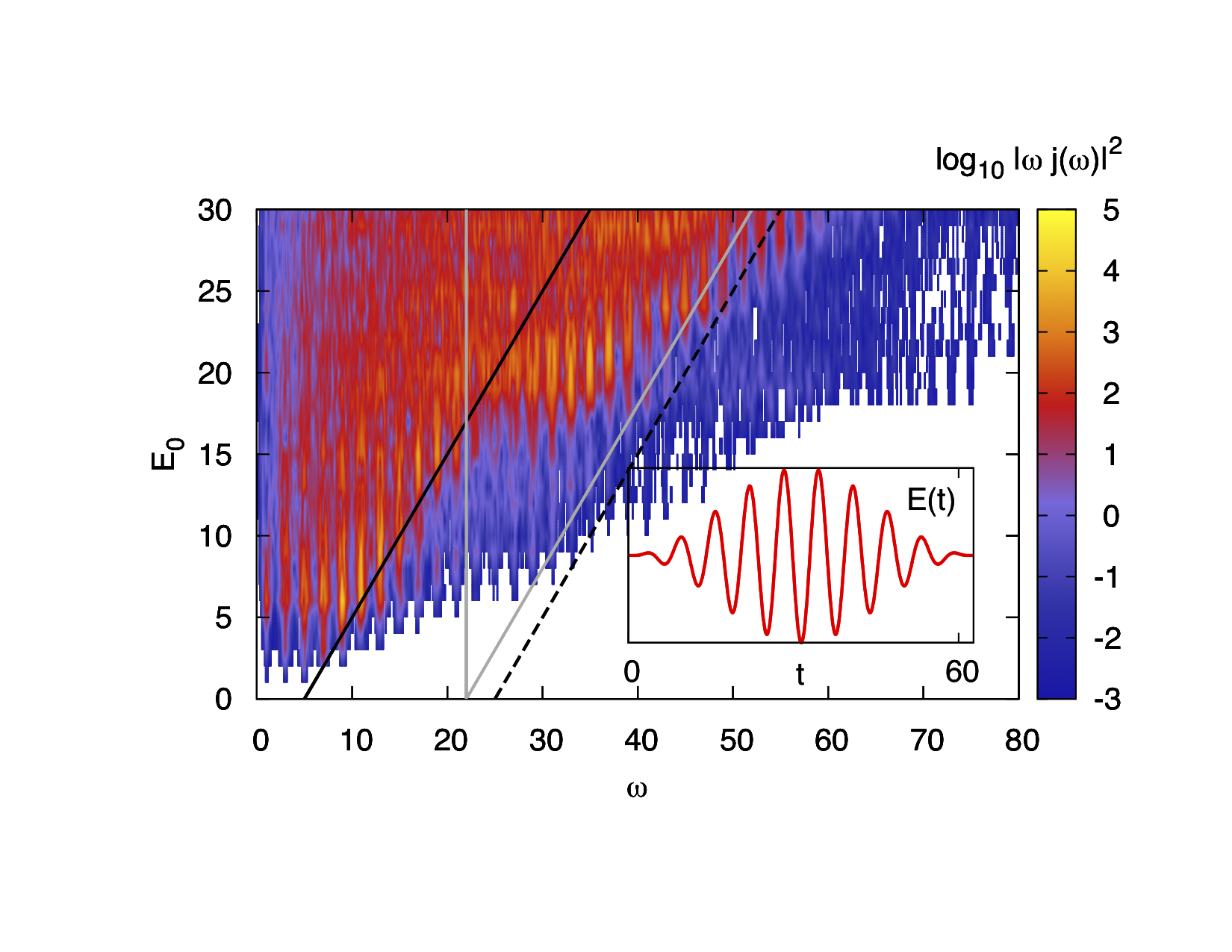} 
   \caption{HHG spectrum of the electron-plasmon model with $\omega_0=17$, $U_\text{scr}=5$, $\lambda=20$ and $\sigma=0$. The black solid line indicates $U_\text{scr}+E_0$, the black dashed line $U_\text{bare}+E_0$, while the gray line indicates $U_\text{scr}+\omega_0+E_0$. The gray vertical line is at $U_\text{scr}+\omega_0$. Inset: Shape of the 10-cycle excitation pulse with central frequency $\Omega=1$ and a $\sin^2$ type envelope.  
   }
   \label{fig:E_vs_w_panel}
\end{figure}

The main panel of Fig.~\ref{fig:E_vs_w_panel} shows the HHG spectra obtained for fixed $U_\text{scr}=5$, $\lambda=20$ and plasmon energy $\omega_0=17$ (corresponding to $g\approx 13$). The solid black line shows the cutoff  $U_{\textrm{scr}}+E_0$, the dashed black line shows $U_{\textrm{bare}}+ E_0$, while the gray line indicates $U_{\textrm{scr}}+\omega_0 + E_0$. 

For strong fields, the high-intensity region of the HHG spectrum exhibits a $U_\text{scr}+\omega_0+E_0$ scaling. There is an almost abrupt change in the cutoff scaling from $U_\text{scr}+E_0$ to $U_\text{scr}+\omega_0+E_0$ near $E_0\approx \omega_0=17$ and one observes a high radiation intensity near $E_0\approx U_\text{scr}+\omega_0=22$. This indicates that the observed crossover to the higher-energy cutoff is triggered by doublon-holon recombinations from nearest neighbor sites, with simultaneous absorption of one plasmon. Expressed in terms of the spectral function (Fig.~\ref{fig:spectral_function_small_w0}), this corresponds to transitions from the first plasmon sideband of the upper Hubbard band to the lower Hubbard band. At field strengths $E_0\gtrsim \omega_0$, and especially around $E_0\approx U_\text{scr}+\omega_0$ the laser field excites plasmons and thus enables these types of recombinations.    

For the present parameters, one cannot easily distinguish a $U_\text{scr}+\omega_0+E_0$ cutoff from a $U_\text{bare}+E_0$ cutoff (see black dashed line). However, a systematic check of different parameter sets confirms that the shift in the cutoff to higher energies is primarily controlled by $\omega_0$, and thus related to plasmon absorption.

\subsubsection{Cut-off behavior for $\omega_0<g$}

\begin{figure}[t]
\centering
\includegraphics[angle=0, width=\columnwidth, width=\columnwidth, bb=90 70 750 530, clip=true]{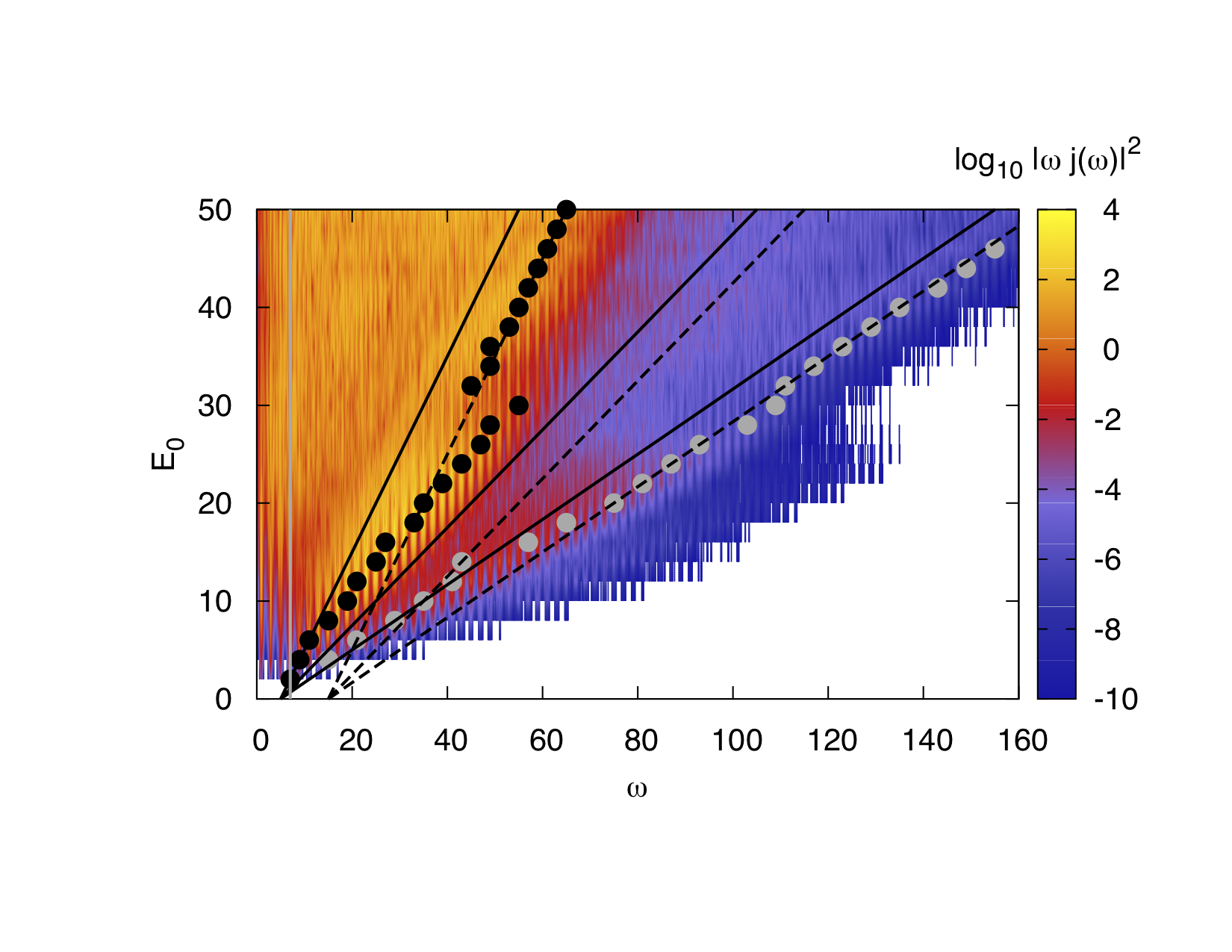}\\
\includegraphics[width=\columnwidth, bb=0 0 385 260, clip=true]{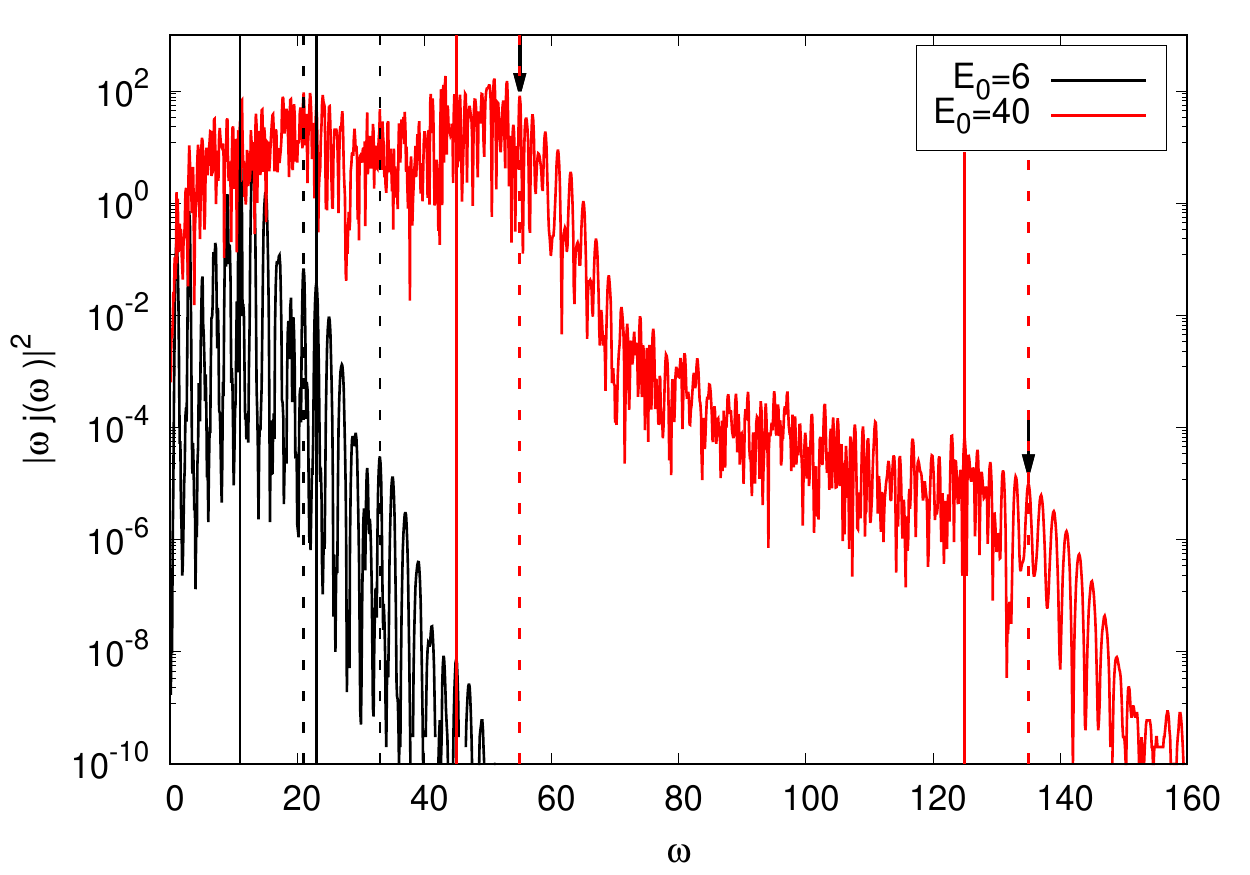}
\caption{Top panel: Harmonic spectra for the Hubbard-Holstein model with $\omega_0=2$, $U_{\textrm{scr}}=5$ and $\lambda=10$. Solid black lines indicate $U_{\textrm{scr}} + nE_0$ and dashed black lines $U_{\textrm{bare}} \pm n E_0$ ($n=1,2,3$). Black dots mark the cutoff energies of the first plateau and gray dots those of the second plateau. 
Bottom panel: HHG spectra at $E_0=6$ and $40$. Here, the solid (dashed) vertical lines indicate $U_{\textrm{scr}} + E_0$ ($U_{\textrm{bare}} + E_0$) as well as $U_{\textrm{scr}} + 3E_0$ ($U_{\textrm{bare}} + 3E_0$). The arrows mark the edges of the plateaus in the $E_0=40$ case.
}   
\label{small_w0_sim}
\end{figure}

In this section, we consider the set of parameters corresponding to the single particle spectral function in Fig.~\ref{fig:spectral_function_small_w0}, which exhibits a relatively dense family of subbands with a dominant peak near $\pm U_\text{bare}/2$ ($\omega_0=2$, $U_{\textrm{scr}}=5$, $\lambda=10$, $g=3.16$). The corresponding HHG spectrum is shown in the top panel of Fig.~\ref{small_w0_sim}. The solid black lines indicate $U_{\textrm{scr}}+E_0$, as well as the higher oder cutoff energies $U_{\textrm{scr}}+nE_0$ ($n>1$). The dashed black lines are the cutoffs associated with $U_{\textrm{bare}}$ ($U_{\textrm{bare}} +E_0$, $U_{\textrm{bare}} +2E_0$ and $U_{\textrm{bare}} +3E_0$).  
The qualitative features of the HHG spectrum are consistent with the naive expectations based on the properties of the single-particle spectral function shown in Fig.~\ref{fig:spectral_function_small_w0}. At low field strength, the cut-off is seen to follow essentially $U_{\textrm{scr}} + E_0$. However, at a sufficiently high field strength, the edge of the dominant HHG plateau crosses over to the $U_{\textrm{bare}}+E_0$ cutoff. Note that the $E_0$ range in this figure is extended compared to the previous one to establish that this crossover is also seen in the third-order cutoff, which switches from $U_{\textrm{scr}} + 3E_0$ to $U_{\textrm{bare}} + 3E_0$. 

Since the identification of the cutoff energies in an intensity plot like Fig.~\ref{small_w0_sim} can be difficult, we also analyzed cuts at fixed $E_0$ to dermine the harmonic orders corresponding to the edges of a plateau, or, in the weak-field regime, to prominent peaks in the HHG spectrum. Examples of such cuts are shown in the lower panel of the figure. Even though the harmonics are not very well defined in the strong field regime, one can identify two plateau-like structures in the HHG signal. The actual cutoffs associated with the first plateau are indicated in the top panel by black dots, and the actual cutoffs associated with the second plateau are indicated by grey dots. The black dots confirm that the dominant plateau indeed exhibits a $U_{\textrm{scr}} + E_0$ scaling for $E_0\lesssim 8$. As one further increases $E_0$, the cutoff energy increases faster than $U_{\textrm{scr}} + E_0$, and eventually even exceeds $U_{\textrm{bare}} + E_0$, which we interpret as a constructive interference between second-neighbor recombination processes in a screened environment and nearest-neighbor processes in an unscreened environment. For $E_0\gtrsim 30$ the cutoff energy of the dominant HHG plateau clearly follows the $U_\text{bare}+E_0$ line. 

Interestingly, in the strong-field regime, the second plateau is associated with next-next-nearest-neighbor recombination processes, as the corresponding cutoff energy crosses over from $U_{\textrm{scr}} + 3E_0$ to $U_{\textrm{bare}} + 3E_0$. No clear plateau-like structure can be associated with the next-nearest-neighbor processes (for $E_0\gtrsim 30$).

Another noteworthy observation is that in the strong-field regime, the plateaus themselves do not exhibit well-defined harmonics, while the regions in between the plateaus, where the intensity decays exponentially as a function of harmonic order, exhibits well defined peaks at odd multiples of $\Omega$. The rather messy HHG signal in the plateau regions may be due to the large number of interfering inter-(side)-band transitions in this system with a relatively small splitting between the multiple boson side bands.

\begin{figure*}[ht]
  \centering
  \subfigure{\includegraphics[width=\columnwidth, bb=90 70 750 530, clip=true]{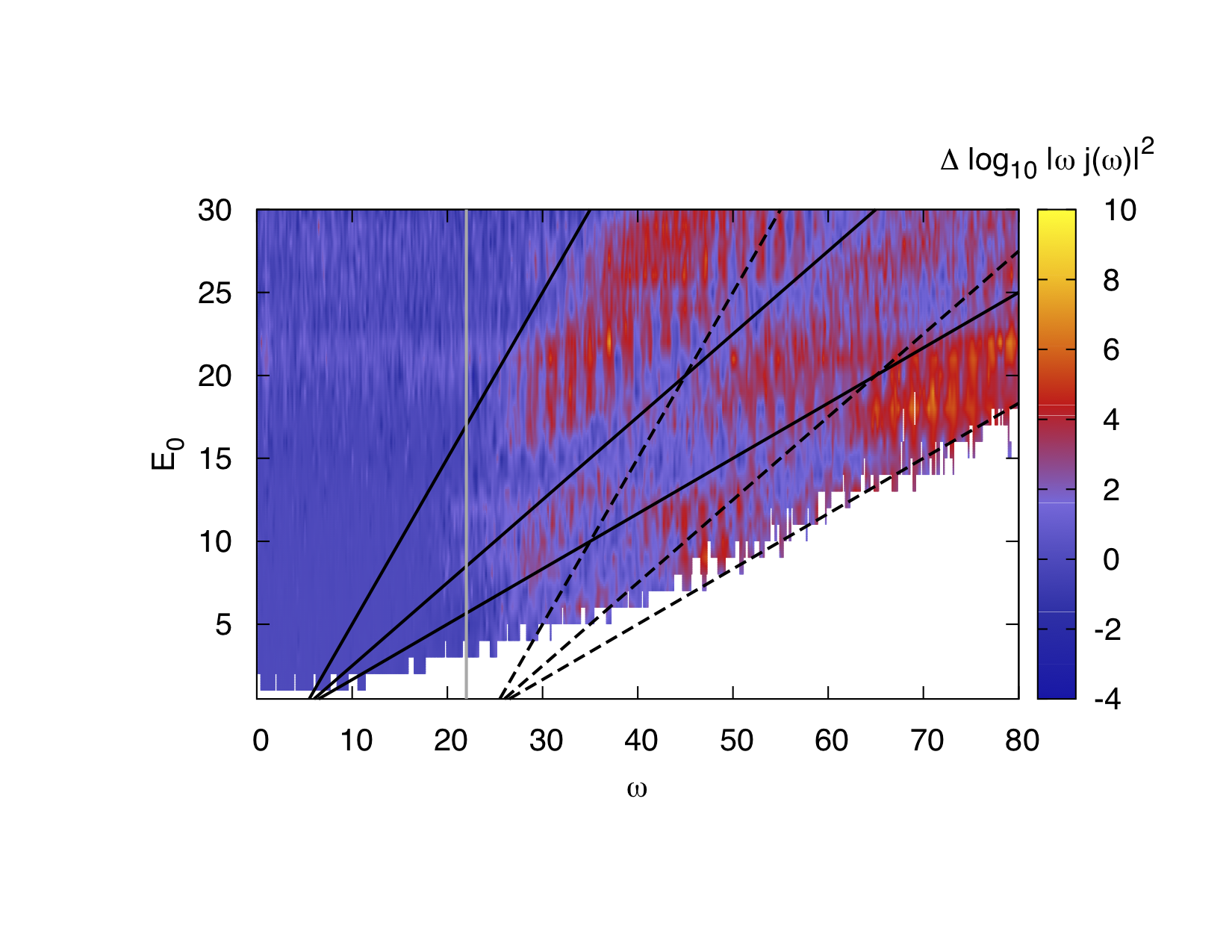}}\hfill
  \subfigure{\includegraphics[width=\columnwidth, bb=90 70 750 530, clip=true]{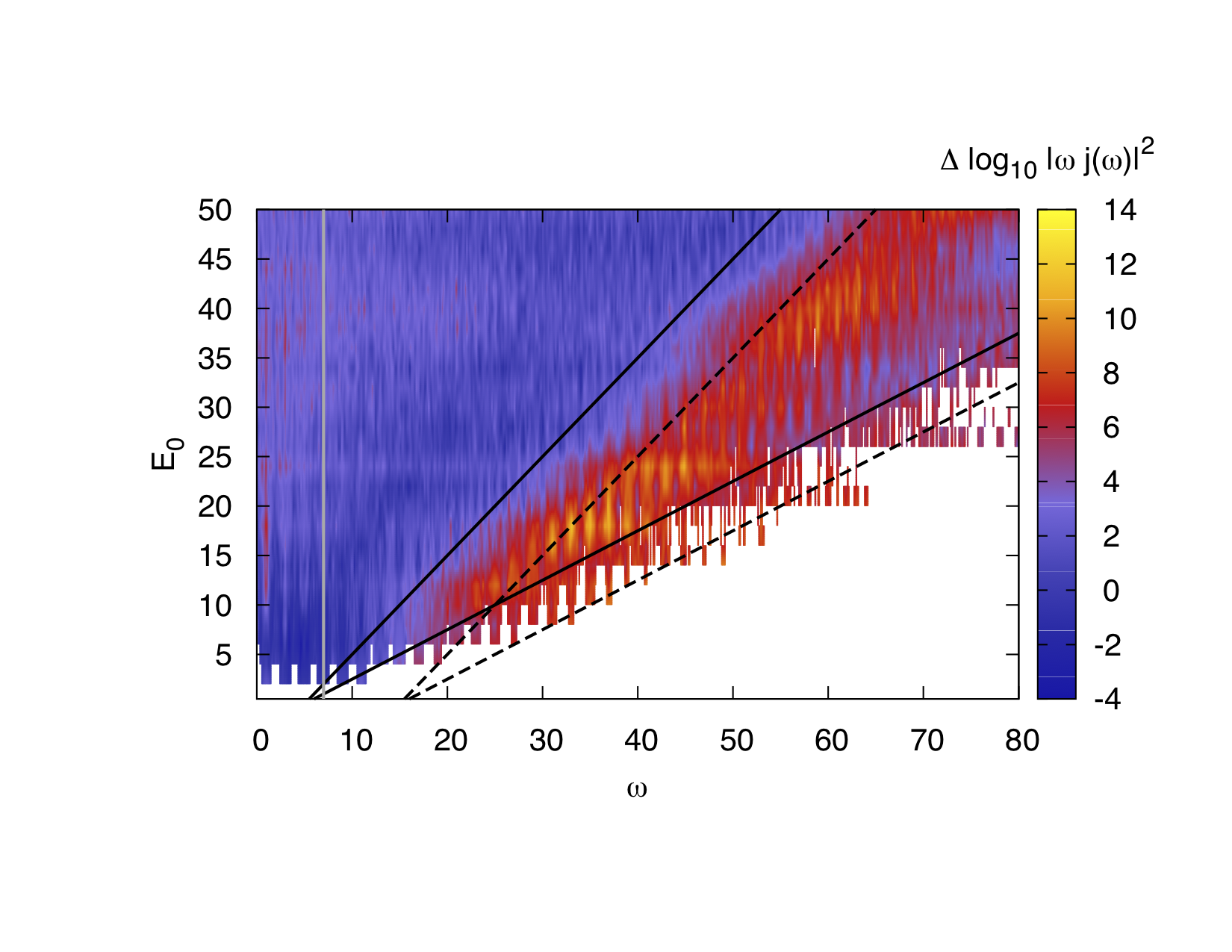}}
   \caption{Left panel: Ratio between the HHG spectra of the Holstein-Hubbard model with $\omega_0=17$, $U_\text{scr}=5$, $\lambda=20$ and the Hubbard model with $U=U_\text{scr}$ and renormalized bandwidth. 
   Right panel: Analogous comparison between the Holstein-Hubbard model with $\omega_0=2$, $U_\text{scr}=5$, 
   $\lambda=10$ with the Hubbard model with $U=U_{\textrm{scr}}$ and renormalized bandwidth. Black solid lines indicate $\omega=U_\text{scr}+nE_0$ and black dashed lines $\omega=U_\text{bare}+nE_0$. The vertical gray lines indicate $\omega_0+U_{\textrm{scr}}$ in both figures. }
    \label{fig:E_vs_w_difference_panel}
\end{figure*}

\subsubsection{Comparison to the Hubbard Model}

As is evident from the pronounced plasmon sidebands in the spectral function (see Fig.~\ref{fig:spectral_function}), 
the electron-boson coupling in the Holstein-Hubbard model introduces additional excitations. It is thus interesting to study the difference in the high-harmonic spectra of the Hubbard and Hubbard-Holstein models. 
For a meaningful comparison between the two models, we choose $U_\text{Hubbard}=U_\text{scr}$ and renormalize the hopping parameter of the Hubbard model to $t_{\textrm{eff}}=t \exp{(-\frac{g^2}{\omega_0^2})}$. This ensures that any difference observed between the two models is not an effect of a modified gap or bandwidth, but the result of the electron-boson interaction.

Figure~\ref{fig:E_vs_w_difference_panel} shows the difference in the logarithms of the high-harmonic intensities, i.e. a log-scale plot of the ratio of the radiation powers $\log[|\omega j_\text{Holstein-Hubbard}(\omega)|^2/|\omega j_\text{Hubbard}(\omega)|^2]$. In the model with $g<\omega_0$ it is found that the electron-plasmon model produces additional harmonics only above $\omega\approx U_\text{scr}+\omega_0$ (gray line) and above the $U_\text{scr}+E_0$ cutoff. This confirms that the additional intensity in the high-energy radiation is associated with transitions between the first plasmon sideband of the upper Hubbard band and the lower Hubbard band, i.e., the absorption of plasmons from the highly excited nonequilibrium system. Note that the complicated structure in the region of high harmonics is at least partially explained by the fact that we divide by the spectrum of the $U_\text{scr}$ Hubbard model, which exhibits plateaus in the HHG spectrum with cutoff eneriges $U_\text{scr}+nE_0$.

The right hand panel shows the results of an analagous comparison for the model with $\omega_0<g$. Here, we notice that additional intensity (compared to the Hubbard model with $U=U_\text{scr}$ and renormalized bandwidth) is observed to the right of the $U_{\textrm{scr}} + E_0$ line, with a large increase for $\omega\gtrsim U_\text{bare}+E_0$. This indicates an important role played by the boson sidebands near $\pm U_\text{bare}/2$ in the single-particle spectrum, which are also the peaks with the dominant weight. The activation of these sidebands in the strong-field regime leads to a shift of the leading HHG cutoff from $U_\text{scr}+E_0$ to $U_\text{bare}+E_0$.

\begin{figure*}[ht]
  \centering
  \subfigure{\includegraphics[width=\columnwidth, width=\columnwidth, bb=90 70 750 530, clip=true]{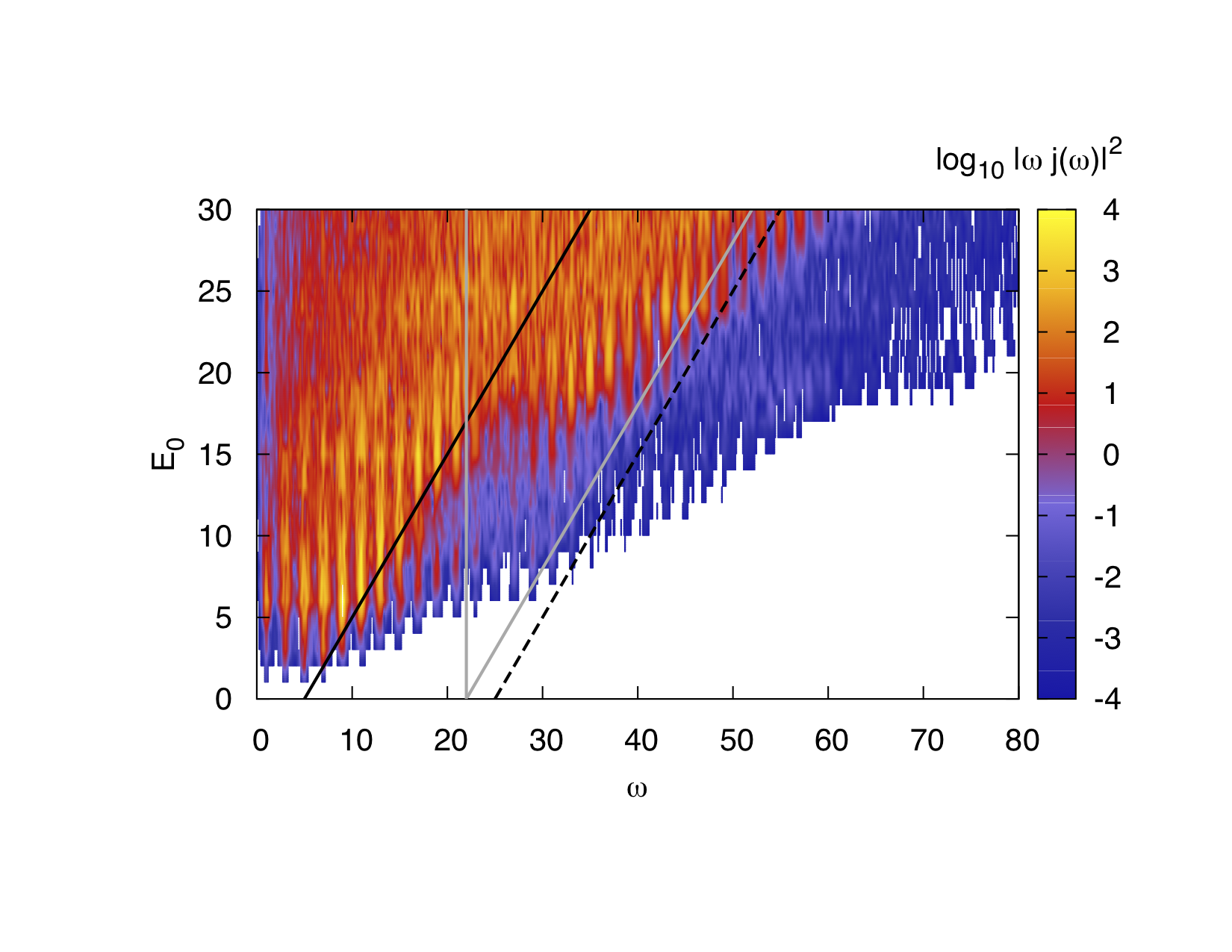}}\quad
  \subfigure{\includegraphics[width=\columnwidth, width=\columnwidth, bb=90 70 750 530, clip=true]{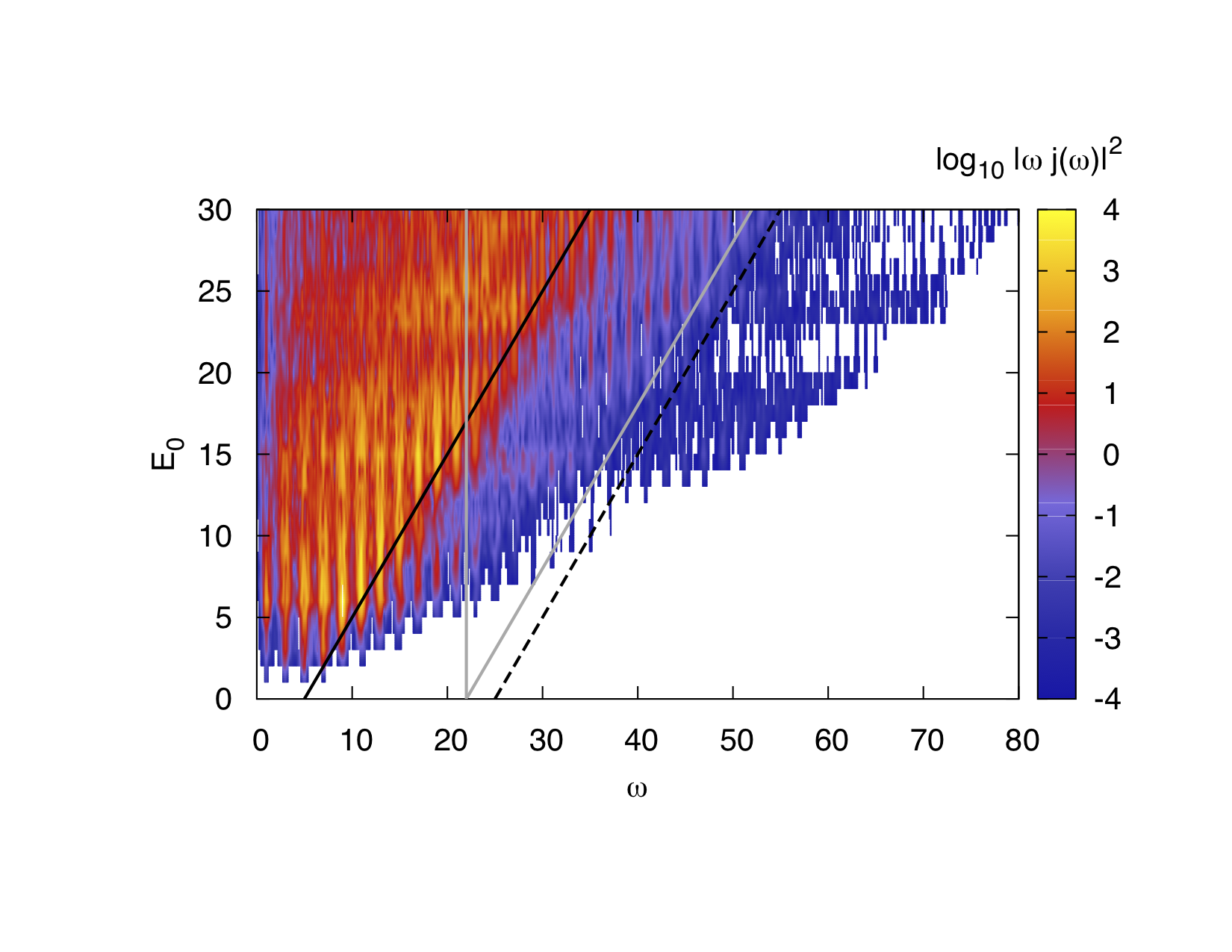}}\quad
  \subfigure{\includegraphics[width=\columnwidth, width=\columnwidth, bb=90 70 750 530, clip=true]{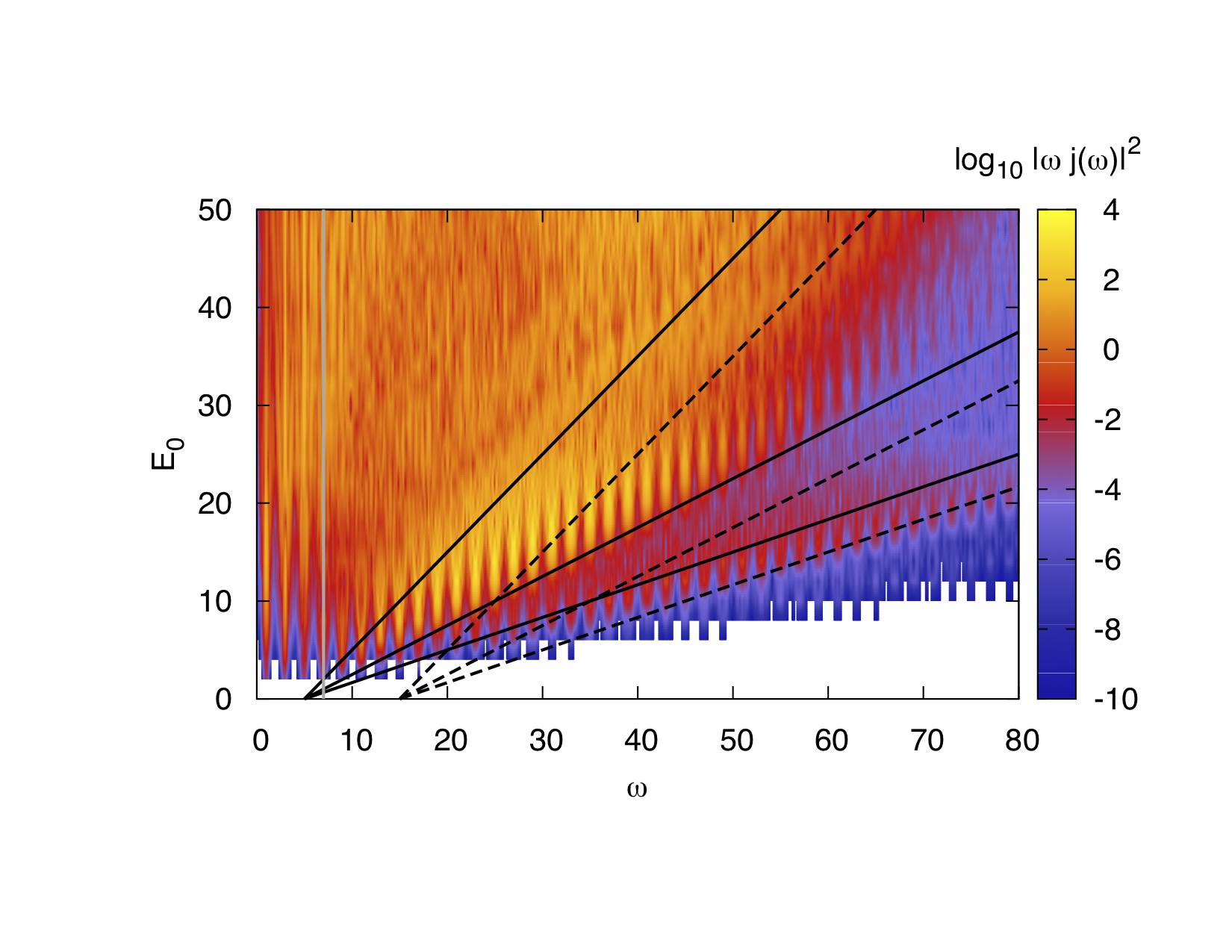}}\quad
  \subfigure{\includegraphics[width=\columnwidth, width=\columnwidth, bb=90 70 750 530, clip=true]{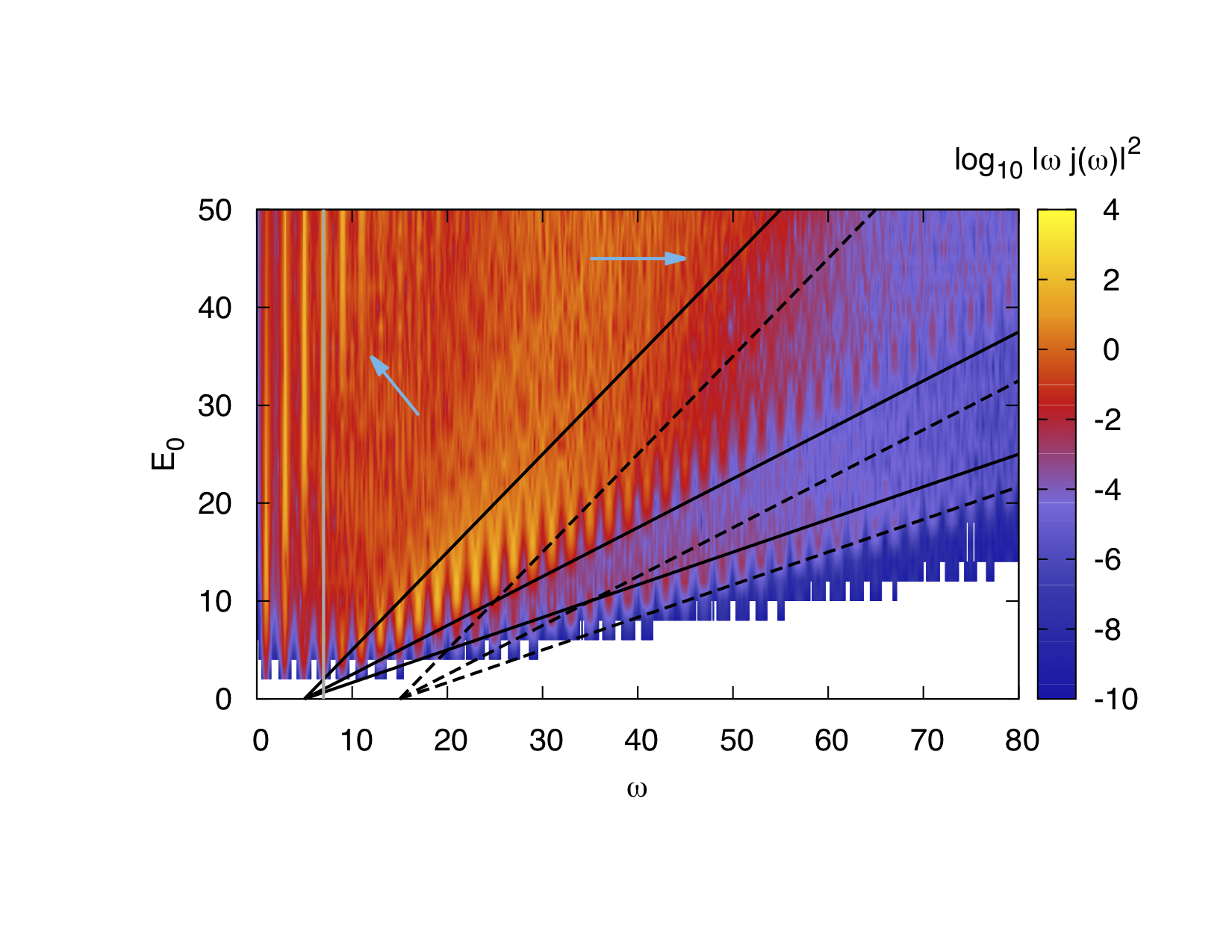}}
  
   \caption{Upper panels: Electron-plasmon model with $\omega_0=17$, $U_{\textrm{scr}}=5$ and $\lambda=20$. Upper left: $\Omega=1$ with $\sigma=0.25$. Upper right:  $\Omega=1$ with $\sigma=1$. Lower panels:  $\omega_0=2$, $U_{\textrm{scr}}=5$ and $\lambda=10$. Lower left: $\Omega=1$ with $\sigma=0.06$.  Lower right: $\Omega=1$ with $\sigma=0.24$. The arrows point out some features not seen in the left panel.    
   }
   \label{fig:many_bosons_panel}
\end{figure*}

\subsubsection{Effects of multiple bosonic modes}

One may note from Fig.~\ref{fig:spectral_function} that in our electron-plasmon models, the effective hopping parameter of the doublons is about $0.5$ (the width of the Hubbard band is about half the bare bandwidth). Accounting also for the field induced renormalization of the bandwidth, we may easily realize a situation in which the driving frequency $\Omega$ exceeds the width of the Hubbard bands. 

In the same figure it is also apparent that in the case of a boson distribution with width $\sigma>0$ the plasmon sidebands get broadened by a factor proportional to $\sigma$, and this broadening persists even in the presence of strong fields. A similar broadening effect and even merging of the sidebands is observed in Fig.~\ref{fig:spectral_function_small_w0}. Hence, depending on the value of $\sigma$ and $\Omega$, intraband excitations within these plasmon sidebands will or will not be possible, and this should manifest itself in the HHG spectra. In particular, we expect that the low-energy harmonics (associated with intraband currents) will be enhanced for $\sigma>0$, while the high-frequency part of the spectrum may exhibit interference effects due to the broad energy distribution of the plasmon-dressed doublons or holes.

In Fig.~\ref{fig:many_bosons_panel} we illustrate the drastic effect of these intraband transitions on the leading energy cutoff. The top panels show HHG spectra for $\Omega=1$ and large boson frequency ($\omega_0>g$).  The left panel is for a boson distribution width $\sigma=0.25$, which is small compared to $\Omega$ and hence we find a crossover from the $U_\text{scr}+E_0$ to $U_\text{scr}+\omega_0+E_0$ cutoff which looks similar to the result found for the single-boson model (Fig.~\ref{fig:E_vs_w_panel}). In the right panel, we show the result for $\sigma=1=\Omega$. Here, the crossover disappears and we only observe the $U_\text{scr}+E_0$ cutoff up to the highest field amplitudes considered. It appears that the intra-band excitations enabled by the broader boson distribution   
wipe out the additional radiation intensity, which in the single-boson case was associated with the activation of inter-band transitions (plasmon absorption). This may be a manifestation of destructive interference between recombination processes with slightly different energies. 

The numerical results for $\omega_0>g$ suggest an approximate condition for the crossover to take place, namely
\begin{equation}
	\sigma \lesssim \Omega.
\end{equation}
Viewing HHG as a spectroscopic method, and assuming a simple form of the dynamically screened interaction as well as a large plasmon energy, this inequality tells us that by varying the driving frequency $\Omega$ and monitoring the crossover behavior, one can deduce the width of the ``plasmon peak" in $\text{Im}U(\omega)$. 

In the case of small boson frequency ($\omega_0<g$), where one observes a crossover from $U_\text{scr}+E_0$ to $U_\text{bare}+E_0$ in the single-boson case, a similar effect of the broadening is observed in the strong-field regime. Here, the crossover is not associated with the activation of transitions between neighboring subbands (single plasmon absorption), but with a field-dependent change in the relative importance of different subbands for the HHG process. As indicated by the left arrow, in the case $\omega_0<g$, a prominent effect of a broadened boson spectrum is an increase in the intensity of the low-energy harmonics, which is consistent with the merging of the side-bands evident in Fig.~\ref{fig:spectral_function_small_w0} and a correspondingly enhanced intra-band contribution to the HHG signal. 

While the crossover to the $U_\text{bare}+E_0$ cutoff at strong fields $E_0$ disappears for larger $\sigma$ (see right arrow), the strong intensity up to approximately $U_\text{scr}+2E_0$ persists at intermediate values of $E_0$. This indicates that these harmonics indeed originate mainly from second nearest neighbor recombination processes in a screened environment.

\subsection{Two-orbital model}
\label{sec:2orbital}

\subsubsection{Half-filled model}

\begin{figure}[ht]
\begin{center}
\includegraphics[angle=-0, width=0.485\columnwidth, bb=50 25 585 650, clip=true]{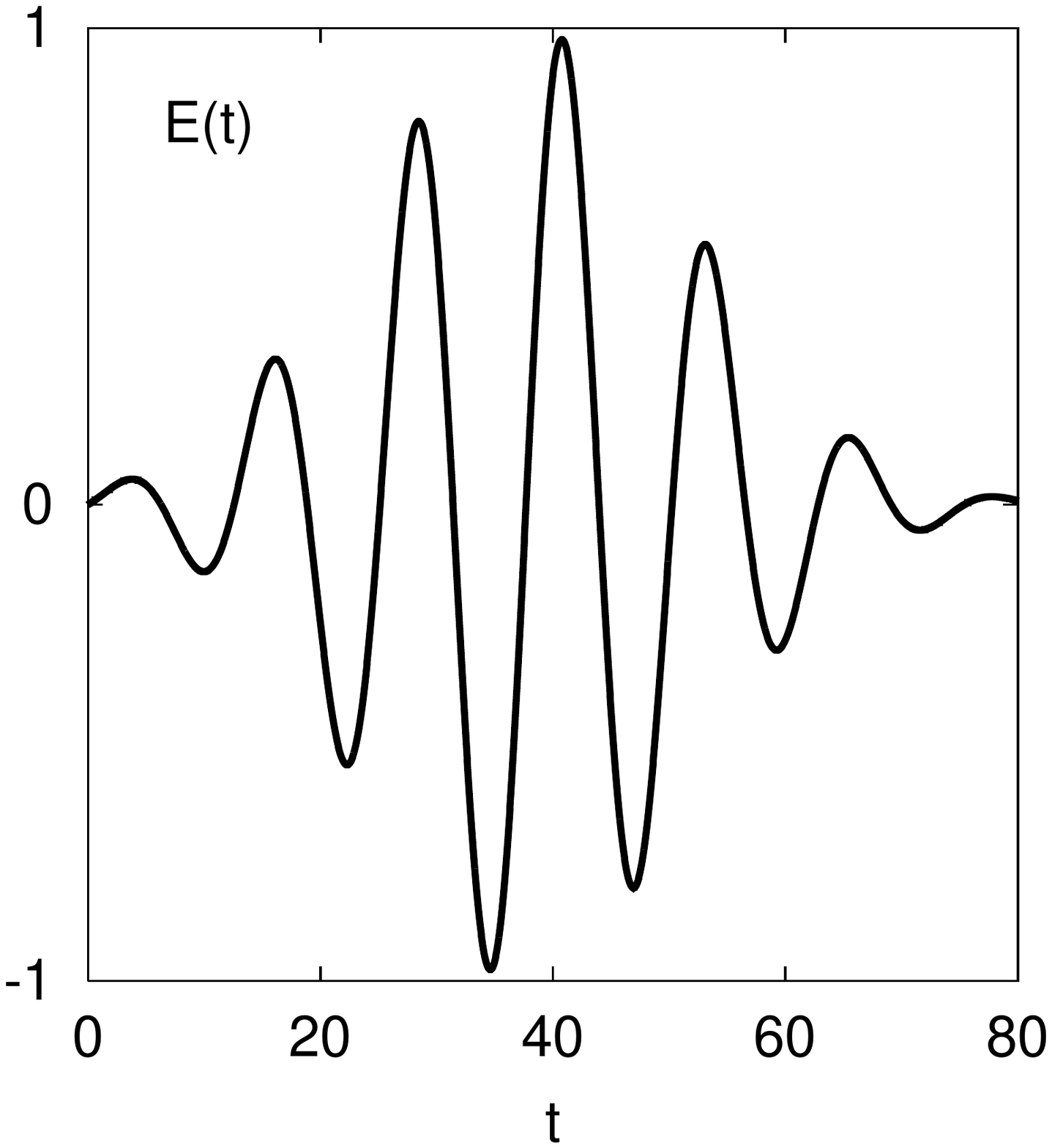}
\includegraphics[angle=-0, width=0.485\columnwidth, bb=50 25 585 650, clip=true]{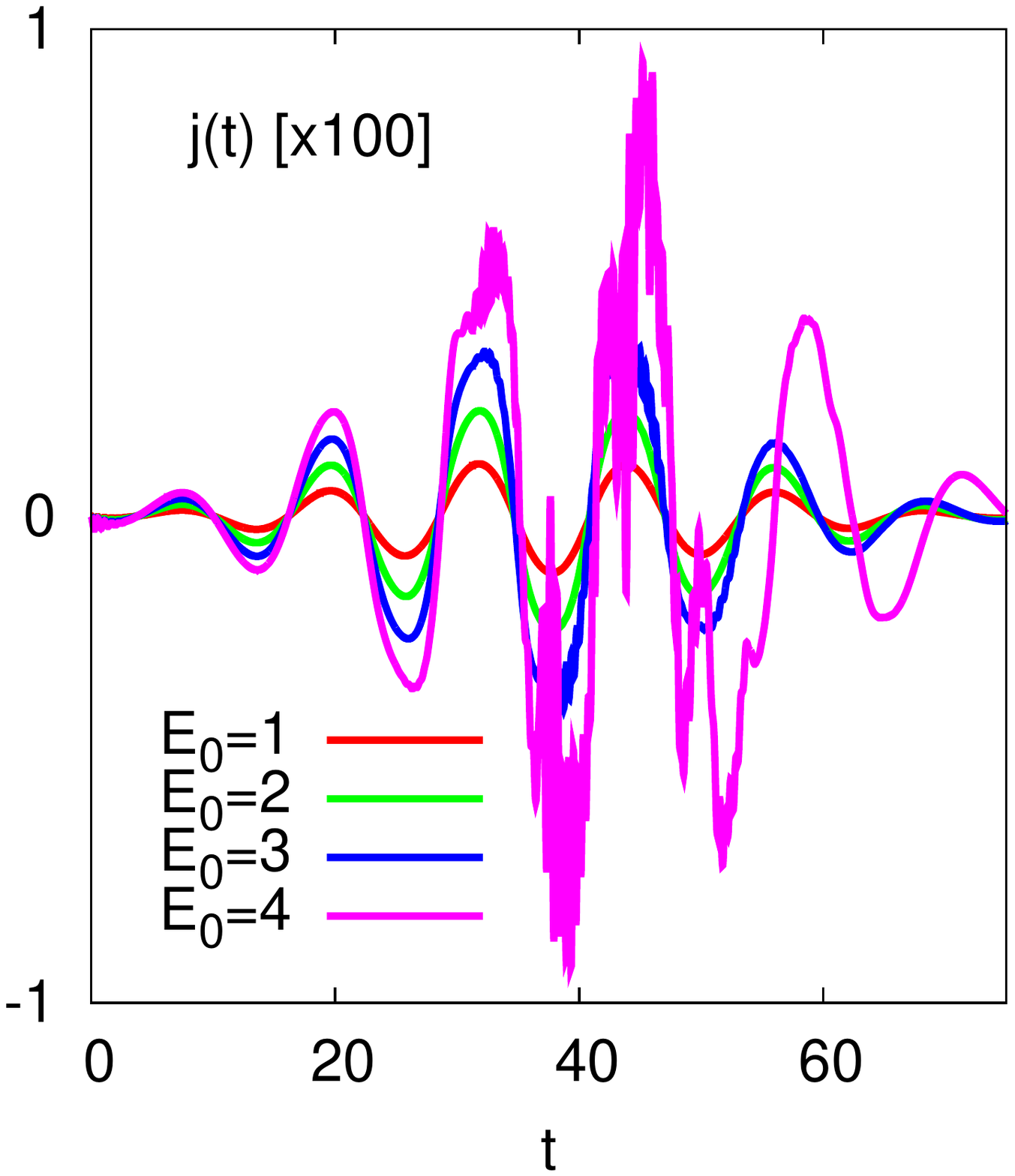}
\caption{Left panel: shape of the few-cycle excitation pulse with frequency $\Omega=0.5$ and amplitude $E_0=1$ used in the two-orbital simulations. Right panel: current for indicated values of $E_0$ ($U=10, J=1$).  
}
\label{fig:pulse}
\end{center}
\end{figure}

\begin{figure*}[ht]
\begin{center}
\includegraphics[angle=-0, width=\columnwidth, bb=90 70 750 530, clip=true]{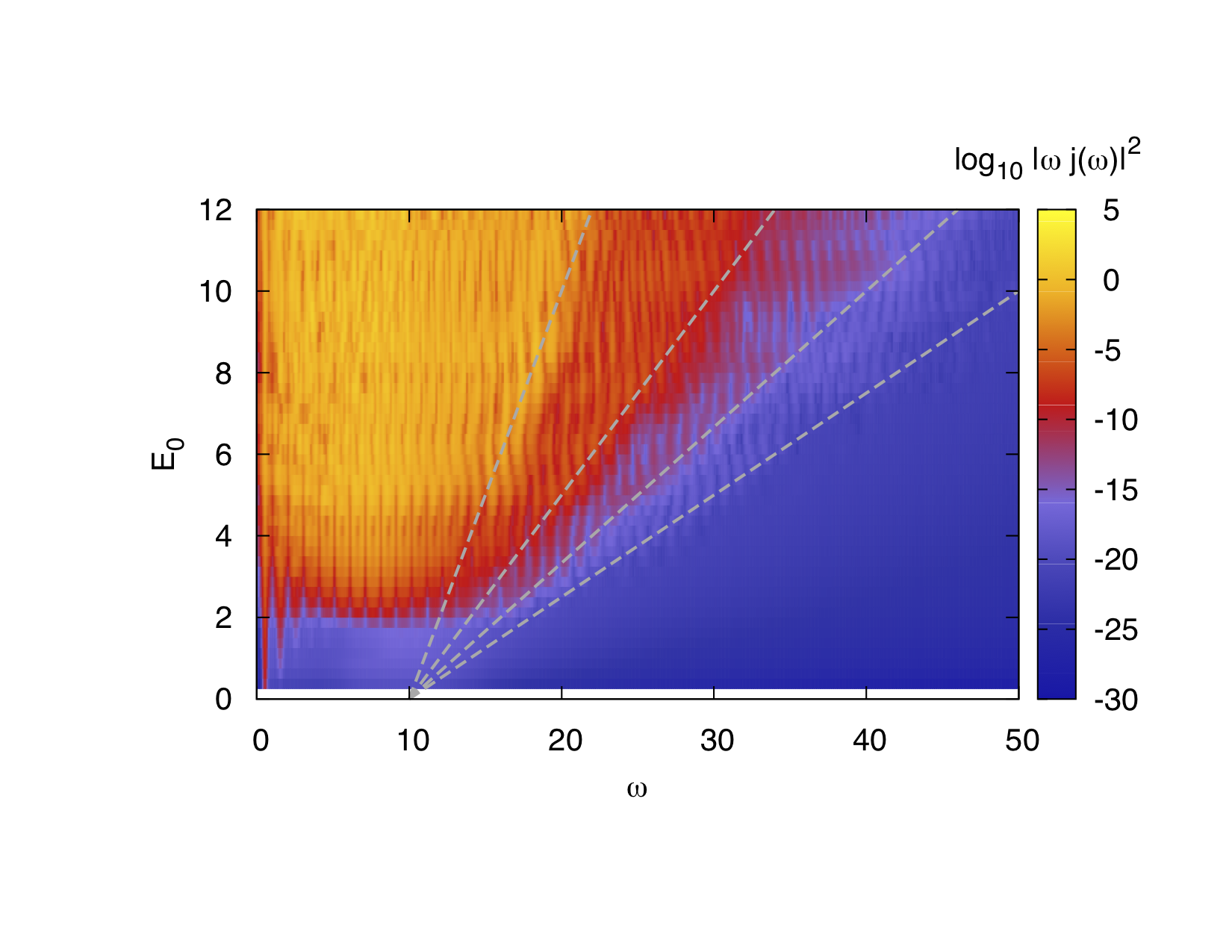}\hfill
\includegraphics[angle=-0, width=\columnwidth, bb=90 70 750 530, clip=true]{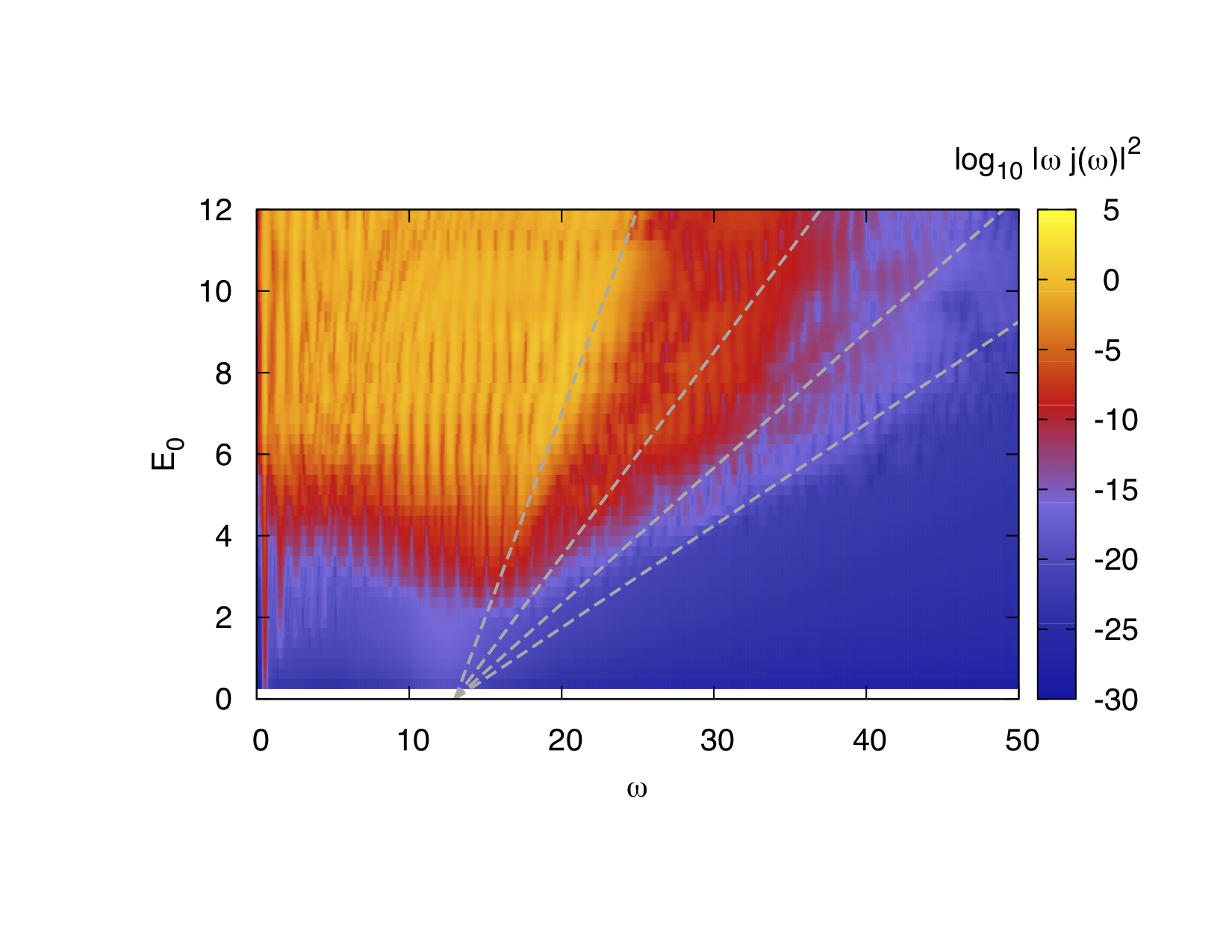}
\caption{High-harmonic spectra $|\omega j(\omega)|^2$ as a function of $E_0$ for $J=0$ (left) and $J=3$ (right). The gray dashed lines represent the cutoffs $U+J\pm n E_0$ ($n=1,2,3,4$). 
}
\label{fig:hhg_line1}
\end{center}
\end{figure*}

In this section we study HHG in the Mott insulating two-orbital model, with the focus on signatures of the Hund coupling. We start with the half-filled system and consider a Mott insulator with a large gap ($U=10$, $\beta=5$), which is driven by a few-cycle electric field pulse with a frequency $\Omega=0.5 \ll \text{gap}$. The form of the pulse with a peak field amplitude $E_0=1$ is plotted in the left panel of Fig.~\ref{fig:pulse}. The current measured for different values of $E_0$ in a system with $J=1$ is shown in the right hand panel. By Fourier transforming these curves, we obtain the HHG spectra $|\omega j(\omega)|^2$ shown for $J=0$ and $J=3$ and a range of field amplitudes in Fig.~\ref{fig:hhg_line1}. 

\begin{figure}[h]
\begin{center}
\includegraphics[angle=-0, width=1.0\columnwidth]{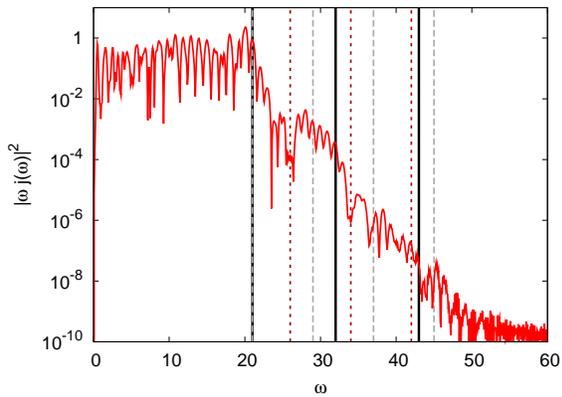}
\caption{High-harmonic spectra for $E_0=8$ and $J=3$. The vertical lines show $U+J+nE_0$ (gray dashed), $U+J+nE_0+(n-1)J$ (solid black), and $2(U+J)+(n-1)E_0$ (dark red dashed), with $n\ge 1$.
}
\label{fig:hhg_cut}
\end{center}
\end{figure}

We are now going to analyze the structures apparent in these HHG spectra, focussing on the strong-field regime. As we mentioned before, in the Mott phase of the single-band Hubbard model, the plateau structures and energy cutoffs can be explained by quasi-local processes: recombination of doublons and holons on $n$th-nearest-neighbor sites.\cite{Murakami_2018a} In a half-filled two-orbital Hubbard model with $J>0$, the half-filled Mott insulator will have predominantly two electrons per site, in a high-spin configuration. An excitation across the Mott gap creates a singlon-triplon pair at an energy cost of $U+J$. If this pair is created on $n$th nearest neighbor sites, the energy released in the recombination process will, in the presence of the oscillating field with strength $E_0$, be in the range $U+J\pm nE_0$ (assuming that no spin-flips occur). The corresponding cutoff values are indicated in  
Fig.~\ref{fig:hhg_line1} by gray dashed lines. They explain some, but not all of the structures. For example, in Fig.~\ref{fig:hhg_cut}, where we plot a cut for $J=3$ at $E_0=8$, it is apparent that these gray dashed lines do not coincide with the edges in the HHG plateaus. 

In the half-filled two-orbital case, the singlon-triplon creation/annihilation may involve local spin (de)exciations. In particular the singlon-triplon creation on $n$th nearest neighbor sites with $n>1$ typically involves string states associated with Hund excitations. An example for $n=2$ is 
$(\downarrow,\downarrow)_{j-1}(\uparrow,\uparrow)_j (\downarrow,\downarrow)_{j+1} \rightarrow (0,\downarrow)_{j-1}(\uparrow\downarrow,\uparrow)_j (\downarrow,\downarrow)_{j+1} \rightarrow (0,\downarrow)_{j-1} (\downarrow,\uparrow)_j (\uparrow\downarrow,\downarrow)_{j+1}$, 
see also the discussion in Sec.~\ref{sec:twoorbitaltheory}. 
The corresponding recombination processes yield cutoff values shifted by multiples of $J$, with the maximum cutoff energy shifted by $(n-1)J$ in the case where $n-1$ sites are flipped back to the high-spin configuration.   
In Fig.~\ref{fig:hhg_cut} we indicate the corresponding maximum cutoffs $U+J+nE_0+(n-1)J$ by solid black lines.  
They show a much better agreement with the measured spectrum, which strongly suggests that the annihilation of string states plays a role in the HHG process in multi-orbital Hubbard systems with Hund coupling. 

\begin{figure*}[t]
\begin{center}
\includegraphics[angle=-0, width=\columnwidth, bb=90 70 750 530, clip=true]{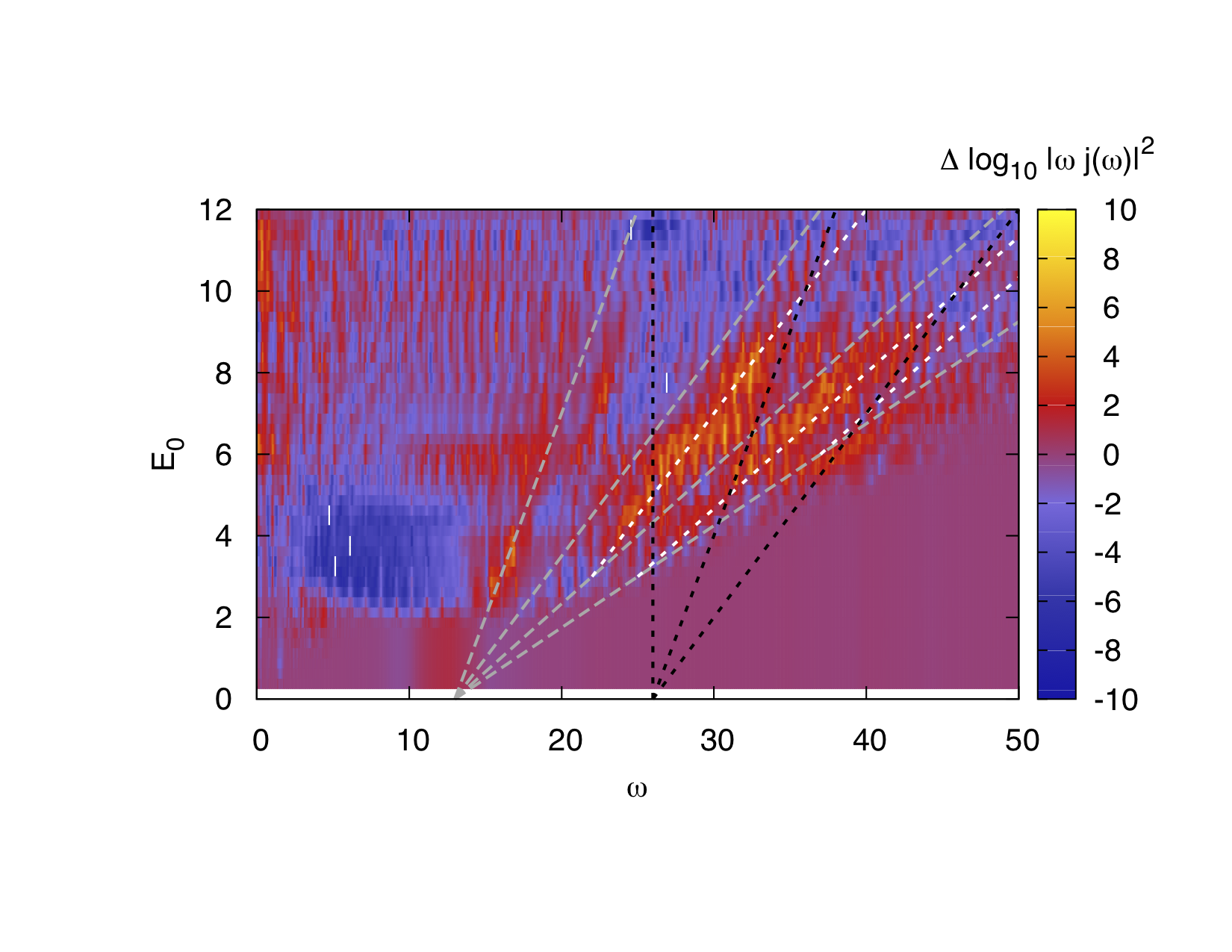}\hfill
\includegraphics[angle=-0, width=\columnwidth, bb=90 70 750 530, clip=true]{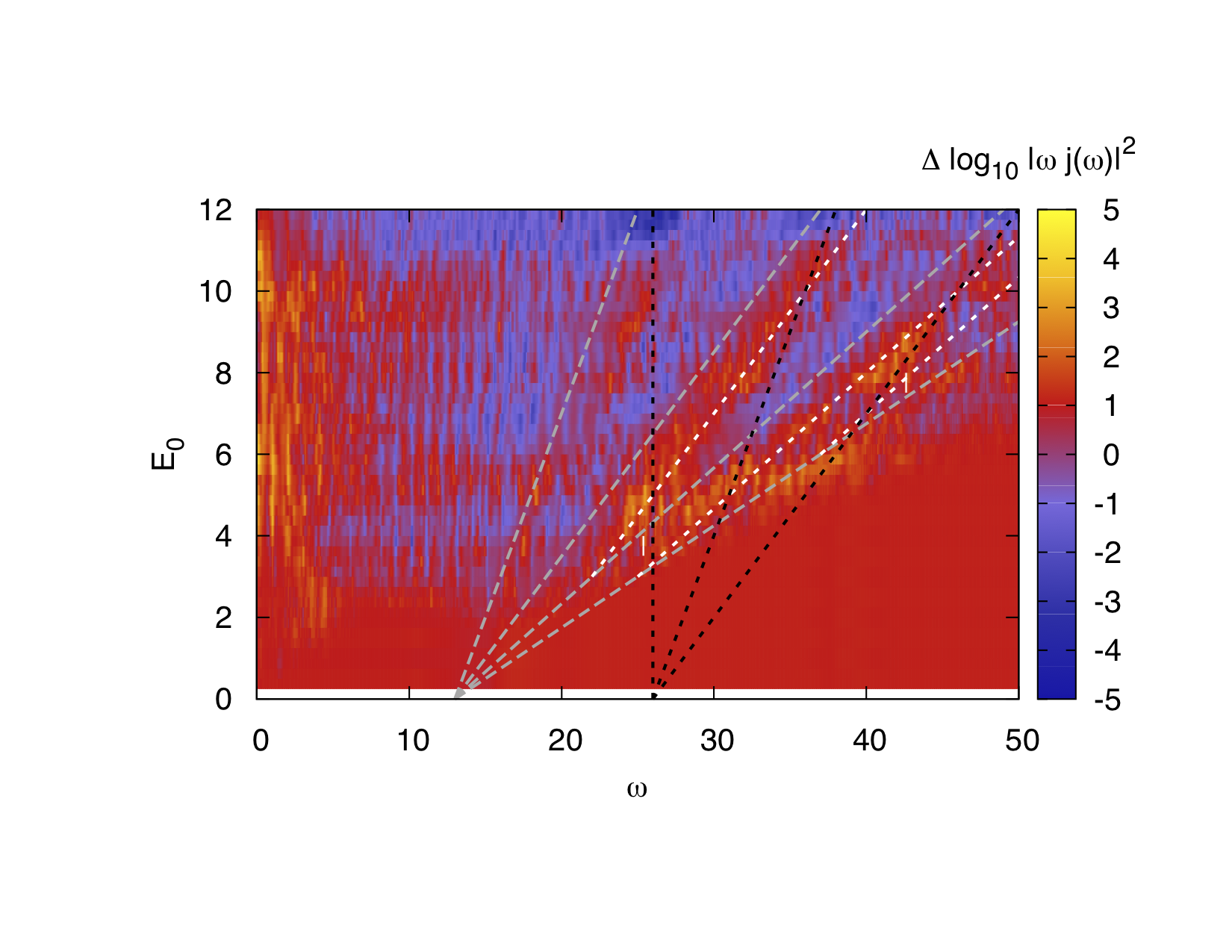}
\caption{The left panel shows a log plot of the ratio of the HHG spectra for the half-filled two-orbital models with $U=13,J=0$ and $U=10,J=3$ (original bandwidth), while the right panel shows the ratio between the HHG spectra for $U=13,J=0$ and  $U=12.6,J=0$ with a bandwidth multiplied by 0.53. The gray dashed lines represent the cutoffs $U+J\pm n E_0$ and the white lines the maximum cutoffs including spin-flip processes, $U+J+n E_0+(n-1)J$. The black dashed lines are the cutoffs associated with second order in $U$ processes.}
\label{fig:hhg_difference}
\end{center}
\end{figure*}

To further analyze this issue, we plot in Fig.~\ref{fig:hhg_difference} the ratio of the HHG spectra for $J=3$ and $J=0$ on a log scale. In the left panel we keep the bandwidths of both models the same ($=4$) and compare $U=10,J=3$ to $U=13,J=0$, so that $U+J$ is identical in both cases. In the right panel, we compare $U=10,J=3,\text{bandwidth}=4$ to $U=12.6,J=0,\text{banwidth}=0.53\cdot 4$. Here the parameters of the $J=0$ system have been chosen such that the Hubbard bands match the positions and widths of the main Hubbard bands at $\omega\approx\frac{U+J}{2}$ in the $U=10,J=3$ spectrum (see pink line in Fig.~\ref{fig:a}). Both figures confirm a shift of the edges of the HHG plateaus to higher energies in the model with Hund coupling. In particular, the cutoff line associated with next-nearest neighbor recombination processes is shifted by $J$, while for next-next-nearest neighbor recombinations, we find both evidence for shifts by $J$ and $2J$ (see white dashed lines which indicate the shifts by $J$ and $2J$, respectively). We also notice that even the $n=1$ line appears to be shifted, at least for large $E_0$, see right panel of Fig.~\ref{fig:hhg_line1} and Fig.~\ref{fig:hhg_difference}. This indicates that nearest-neighbor (in the direction of the field) recombination processes with simultaneous Hund de-excitation via hopping perpendicular to the field play a role in the HHG.

The complex structures evident in Fig.~\ref{fig:hhg_line1} suggest interference effects from other types of processes. At energies $\gtrsim 2U+2J$ one can expect to see higher order processes like  $(\uparrow,\uparrow)_i (\downarrow,\downarrow)_j (\uparrow,\uparrow)_k (\downarrow,\downarrow)_l \rightarrow (0,\uparrow)_i (\uparrow\downarrow,\downarrow)_j(\uparrow,0)_k (\downarrow,\uparrow\downarrow)_l$. If the production of the two singlon-triplon pairs occurs in the direction of the field, the maximum cutoff value associated with the recombination is $2(U+J)+2E_0$. However, the second singlon-triplon pair could also be excited in a direction perpendicular to the field in which case the cutoff becomes $2(U+J)+E_0$. If it occurs against the field we expect a cutoff $2(U+J)$. The corresponding cutoffs energies are indicated in Fig.~\ref{fig:hhg_difference} by the dashed black lines and explain some of the intensity modulations seen at high $\omega$. 
In Fig.~\ref{fig:hhg_cut}, which shows a cut at $E_0=8$, the corresponding cutoff energies are indicated by the dark red dashed lines. These confirm a (possibly negative) interference effect between first order and second order processes. 
It thus appears that in strongly interacting Mott systems, in contrast to semi-conductors, higher-order interaction processes also play a role in the high-energy region of the HHG spectrum and the associated currents interfere with those of the leading order processes.

\subsubsection{Quarter-filled model}

As illustrated in Fig.~\ref{fig:a}, adding an electron to a singly occupied site creates doublon states with energies $U-3J$, $U-2J$ or $U$, so that the upper Hubbard band for large $J$ splits into three subbands. In the quarter filled model, a charge excitation across the gap corresponds to the creation of an empty site and a doubly occupied site. Doublons moving in the background of singly occupied states do not leave behind a string of Hund excitations, because the flipping of the spin does not cost any energy. We thus do not expect to see signatures of string annihilation in the cutoff energies of the high harmonic spectra. On the other hand, the presence of three Hubbard subbands will lead to a rich cutoff structure, which reveals the Hund energy $J$.

\begin{figure*}[t]
\begin{center}
\includegraphics[angle=-0, width=\columnwidth, bb=90 70 750 530, clip=true]{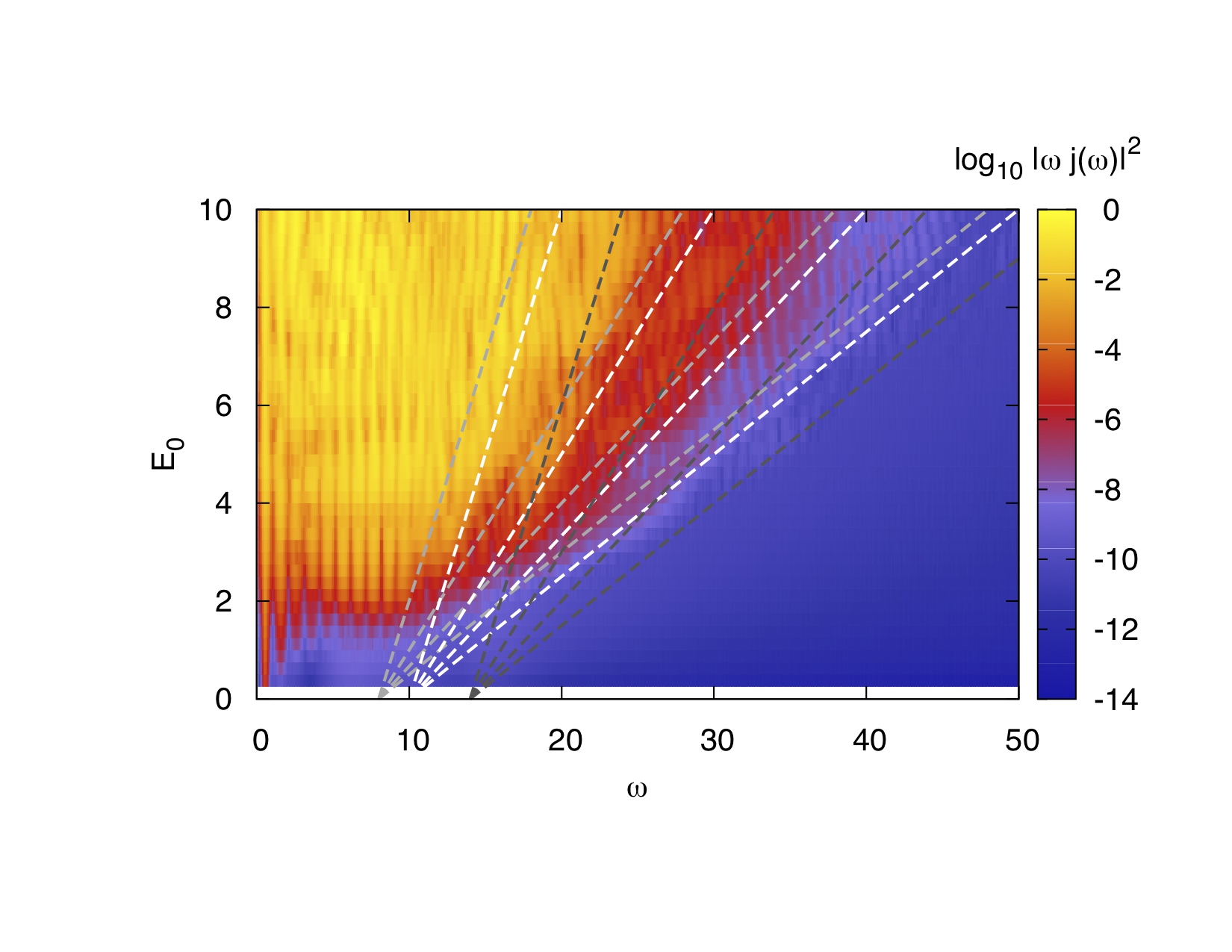}\hfill
\includegraphics[angle=-0, width=\columnwidth, bb=90 70 750 530, clip=true]{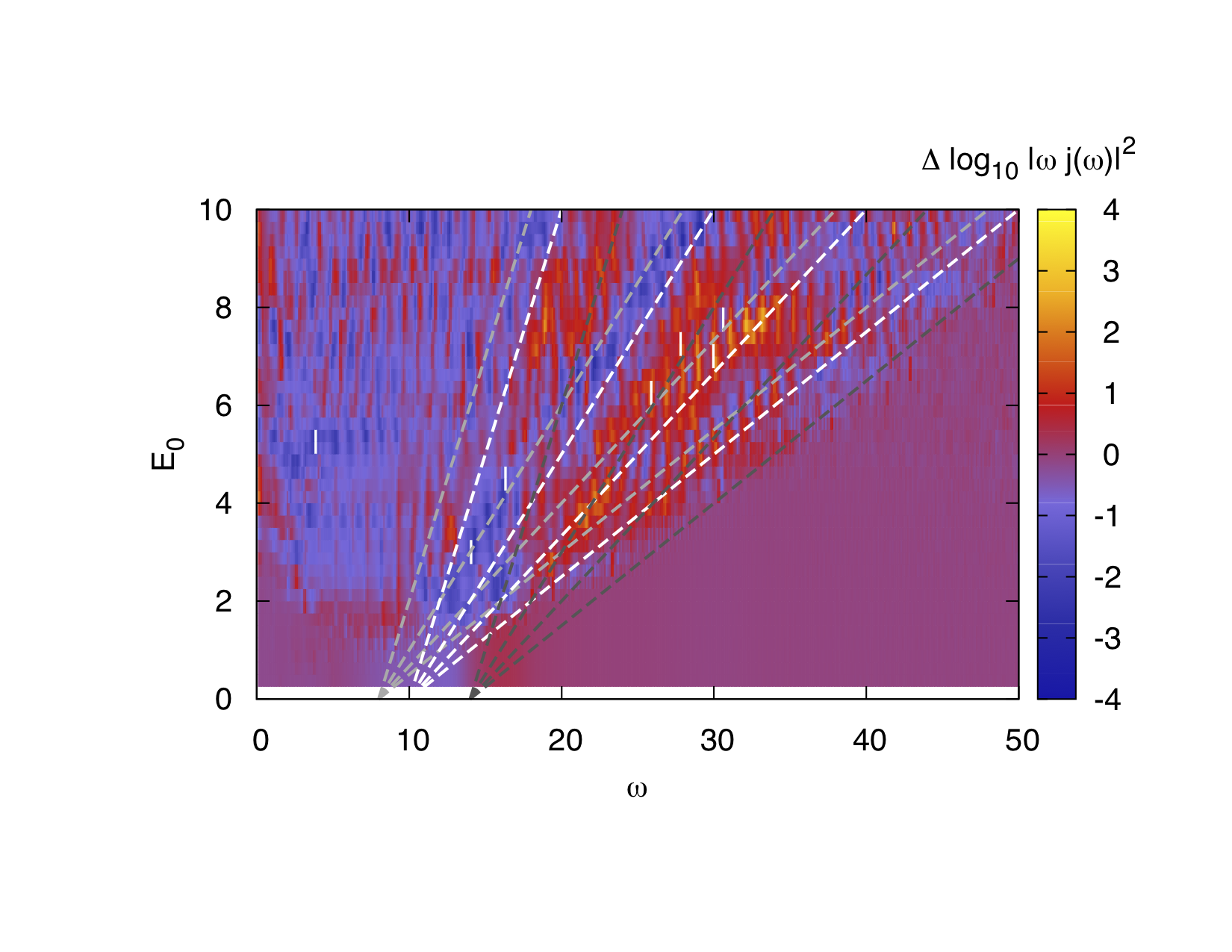}
\caption{HHG spectra of the quarter filled two-orbital model. The left panel shows the result for $U=14$, $J=2$ and the right panel the difference of this spectrum to the result for $U=9$, $J=0$. Light gray dashed lines show the cutoffs $U-3J+nE_0$, white dashed lines $U-2J+nE_0$ and dark gray dashed lines $U+nE_0$ (for $U=14$, $J=2$).
}
\label{fig:hhg_quarter}
\end{center}
\end{figure*}

In the left panel of Fig.~\ref{fig:hhg_quarter}, we plot the HHG spectrum for $U=14$, $J=2$. The light gray dashed lines indicate the cutoff energies $U-3J+nE_0$, the white dashed lines the cutoffs $U-2J+nE_0$ and the dark gray dashed lines the cutoffs $U+nE_0$. While the cutoffs associated with $U-3J$ and $U-2J$ are hard to distinguish, since the corresponding Hubbard subbands are not clearly separated for $J=2$ (see Fig.~\ref{fig:a}), the cutoffs associated with $U$ can be clearly identified.

In the right panel, we show a log plot of the ratio between the $U=14,J=2$ HHG spectrum and the result for a model with $U=9,J=0$. The $U$ value in the latter case has been chosen in between $14-3\cdot 2$ and $14-2\cdot 2$, so that the upper Hubbard band of the $J=0$ model covers approximately the same energy range as the two lower subbands in the $J=2$ model. Near the $U-3J+nE_0$ and $U-2J+nE_0$ cutoff lines, we expect a reduced intensity in the $J=2$ case, because of the reduced spectral weight compared to the $J=0$ case (see black and blue curves in the right panel of Fig.~\ref{fig:a}), while in the energy region between the $U-2J+nE_0$ and $U+nE_0$ cutoff lines we expect an enhancement. This is indeed what is found in the right hand panel of Fig.~\ref{fig:hhg_quarter}, where an increase in intensity is mainly observed between the white dashed and dark-gray dashed lines.

\section{Discussion and conclusions}
\label{sec:conclusions}

We have analyzed the high-harmonic spectra of two types of Mott insulating Hubbard-type systems whose dynamics is influenced by bosonic excitations. The results for the electron-plasmon system, described by a Hubbard-Holstein model with large boson frequency, revealed a crossover between two different cut-off laws. 
In the weak-field regime the cutoff scales as $U_\text{scr}+E_0$, while in the strong-field regime the high harmonics plateau extends up to $U_\text{scr}+\omega_0+E_0$ (for $\omega_0>g$) or $U_\text{bare}+E_0$ (for $\omega_0<g$). Similar crossovers can also be observed in the cutoff energies of the higher-order plateaus associated with $n$th ($n>1$) nearerst neighbor recombination processes. 
The two different crossover behaviors for $\omega_0 < g$ and $\omega_0> g$ were further supported by 
comparing the Holstein-Hubbard model results to those for a Hubbard model with effectively renormalized bandwidth and $U=U_\text{scr}$. In the case of $\omega_0>g$, the additional intensity in the high-energy radiation contribution can be associated with the absorption of plasmons which are excited at field strengths $E_0\gtrsim \omega_0$. 
In the $\omega_0<g$ case, a different picture emerged.  
Here, the additional radiation power was observed for energies $\gtrsim U_\text{bare}+E_0$. In this coupling regime it is not the transitions between the relatively tightly spaced sidebands which drive the crossover, but the gradual shift in the relative importance of different sidebands for the HHG process. At low field strength the peaks in the single particle spectrum near $\pm U_\text{scr}/2$, which define the Mott gap, play a prominent role, while for larger $E_0$, the peaks near $\pm U_\text{bare}/2$, which have the largest weight, become more relevant.  

If the width of the plasmon peak in $\text{Im}U(\omega)$ is larger than the driving frequency, intraband excitations are activated within the plasmon sidebands. This affects the population within these sidebands and the energies which can be released by interband transitions, resulting in a destructive interference between the plasmon assisted processes. In  systems with $\omega_0>g$, where interband transitions underpin the crossover behavior, one thus observes the disappearance of the crossover and a leading cutoff which scales as $U_\text{scr}+E_0$ up to large field strengths. 
In this regime, the sensitivity of the HHG spectrum on the width of the plasmon peak allows in principle to extract this quantity by tracking the crossover behavior as a function of driving frequency $\Omega$. Hence, by studying HHG spectra for a broad range of $E_0$ it is in principle possible to determine whether $g<\omega_0$ or $g>\omega_0$, and in the former case the value of the screened interaction and of $\omega_0$ can be extracted.

The second model which we considered was a two-orbital model with Hund coupling. In the half-filled system, local spin excitations have a strong effect on the dynamics of charge carriers.\cite{Strand_2017} Singlons and triplons moving in the background of predominantly high-spin doublon sites can leave behind a string of low-spin states (Hund excitations), thereby transferring kinetic energy into potential energy. In the presence of a strong periodic driving field, the energy stored in these strings (a large energy of order $nJ$, where $n$ is the length of the string) can be released upon recombination of the singlon and triplon. This results in high harmonic plateaus which extend to higher energies than what would be expected from the splitting of the Hubbard bands. We have demonstrated this effect by comparing spectra for models with and without Hund coupling and appropriately adjusted gap size and bandwidth. In the quarter filled model, the strings of low spin states are absent, but the Hubbard bands split into subbands, which for large enough $J$ results in three separate cutoffs, associated with the three different types of doublon states that can be produced by photo-excitation. As in the case of the electron-plasmon model, we found that all the relevant energy scales of the atomic problem (here $U$ and $J$) are reflected in the field dependence of the spectrum so that a careful analysis of high harmonic spectra of correlated multi-band materials may give access to these parameters, which are of crucial importance for the theoretical modelling and hard to obtain from ab-initio calculations.

\acknowledgements

We thank M. Eckstein for helpful discussions. The calculations have been performed on the Beo04 and Beo05 clusters at the University of Fribourg, using a software library developed by M. Eckstein and H. Strand. This work has been supported by the European Research Council through ERC Consolidator Grant No. 724103, and by the Swiss National Science Foundation through Grant No. 200021\_165539. PW acknowledges the hospitality of the Aspen Center for Physics during the Summer 2019 program. 

\bibliographystyle{apsrev4-1}
\bibliography{bibliography.bib}

\end{document}